\documentclass[aps, pra, preprint, amsmath]{revtex4-1}

\usepackage[T1]{fontenc}
\usepackage[utf8]{inputenc}
\usepackage{amsmath, amssymb, amsfonts, bm, amsthm, braket}
\usepackage{rotating, graphicx}
\usepackage{color}
\usepackage{hyperref} 

\newcommand{\half}{\frac{1}{2}}

\DeclareMathOperator{\Tr}{Tr}

\newcommand{\Hilb}{\mathcal{H}}

\begin{document}
\date{\today}
\flushbottom

\title{Noise-Induced Coherence in Molecular Processes}
\author{Amro Dodin}
\affiliation{Department of Chemistry, Massachusetts Institute of Technology, Cambridge, Massachusetts, 02139, USA}
\author{Paul Brumer}
\affiliation{Chemical Physics Theory Group, Department of Chemistry, and Center for Quantum Information and Quantum Control, University of Toronto, Toronto, Ontario M5S 3H6, Canada}

\maketitle

\section{Abstract}
\label{sec:abstract}
Excitation of molecules by 
incident incoherent electromagnetic radiation, such as sunlight,
is described in detail and contrasted with the effect of
coherent (e.g. laser) light. The nature of the quantum coherences induced
by the former, relevant to transport processes in nature and in
technology, is emphasized. Both equilibrium and steady state scenarios 
are discussed, Three examples: simple models, calcium 
excitation in polarized light, and the isomerization of retinal
in rhodopsin are used to expose the underlying qualitative nature
of the established coherences.

\newpage

\section{Introduction}
\label{sec:intro}
Understanding how electromagnetic radiation drives the electronic and nuclear dynamics of quantum systems is one of the central questions of atomic, molecular and optical physics.
Over the past century, a large body of theoretical and experimental work has been developed for understanding laser driven excitations, leading \textit{inter alia} to the  development of powerful nonlinear spectroscopies for probing dynamics on short timescales and to a variety of technologies that harness quantum optical properties.
At the heart of these discoveries is an understanding of how an interaction with a laser field, which has deterministic and well understood temporal properties, encodes quantum phase in the material system.
However, in recent years, efforts to understand the dynamics of quantum heat engines and of biomolecular systems under realistic conditions has sparked a growing interest in generalizing these insights to systems driven by noisy stochastic EM fields.
These studies of noise-driven dynamics have revealed new mechanisms through which noisy fields can encode quantum phase in material systems that differ significantly from those seen in laser-driven dynamics.
The resulting excitation phase can modify the photophysical properties of the system in macroscopically observable ways, for example by modifying the quantum yield of photo-induced molecular processes, or the spatial profile of atomic fluorescence.
These noise-induced phenomena have attracted significant interest due to their potential role in natural biomolecular processes (e.g.,  photosynthetic light harvesting or human vision), quantum heat engines and bio-inspired photovoltaics.

To define the types of problems that we focus upon in this overview, consider a molecule driven by various types of light.
As a first example, the molecule may be driven by a continuous wave laser represented by a sinusoidal electric field with a well-defined frequency and phase.
This common excitation scenario will predominantly excite molecular transitions with an energy in resonance with the driving field.
The resulting excited state dynamics will therefore only reflect the behavior of those specific states at that specific energy.
At the opposite  (but still coherent) extreme, one may consider excitation by an infinitesimally short laser pulse, represented by a very narrowly peaked electric field in time (e.g.,  a Dirac $\delta$ function).
Such an excitation will, instead, generate a wave-packet in the excited electronic state that exactly resembles the shape of the ground-state wavepacket.
The resulting dynamics will then reflect the behavior of a wave-packet with that specific spatial profile generated at that specific time.
In both of these cases, it is the well-defined spectral and temporal characteristics of the exciting field that allows us to restrict attention to a specific excited state wavefunction.
The specificity of these laser driven excitations is precisely what has made them useful for a wide variety of experimental and technological tasks.
From a molecular perspective, we can say that deterministic characteristics of the exciting field are encoded in the quantum phases of the generated superposition.

By contrast, consider a molecule driven by sunlight, which is well-described as a white-noise electric field.
This stochastic field does not show any of the well-defined spectral and temporal characteristics of coherent light fields and therefore will generate a wide variety of different stochastic excitations that must be averaged over to determine the system behavior.
Clearly, the system behavior under such an excitation will differ significantly from that resulting from  the deterministic laser driving.
Remarkably, however, even when the field is completely stochastic and uncorrelated, the response of the system preferentially encodes a specific excited state.
The quantum phase of this excited state superposition is referred to as \textit{noise-induced coherence}.

In this tutorial we  outline the theoretical tools and concepts used to understand noise-driven dynamics, and highlight their practical implications in atomic and molecular systems.
In section \ref{sec:rho}, we begin by summarizing the density matrix description of quantum states required to treat stochastically driven systems, and show how the quantum phase between states is encoded in the off-diagonal density matrix elements, known as  coherences.
Section \ref{sec:excit} then presents a time-domain picture for the dynamics generated by coherent and incoherent light in terms of interfering pathways through energy eigenstates.
This interfering pathways picture allows us to contrast how coherent laser light and incoherent light encode quantum phase in atomic and molecular systems.
Section \ref{sec:coh} uses this time-domain formalism to explain how noisy fields can generate coherences and to describe the competing processes at play in realistic scenarios.
We then consider the steady-states that dominate the properties of these noise-driven systems in Section \ref{sec:transport}.

This analysis also allows us to describe how local detailed-balance conditions guarantee the emergence of a thermodynamic steady state that typically shows no coherences between energy eigenstates and explain how interaction with two baths at different temperatures breaks these detailed balance conditions and allows for the generation of a non-equilibrium steady state with coherences between energy eigenstates.
These steady-state coherences are associated with, e.g., the transport of energy though the system from a high temperature bath to a low temperature bath and are therefore referred to as \textit{transport-induced coherences}.
In Section \ref{sec:examples} we describe the role played by these coherences in realistic atomic and molecular systems by considering the excitation of atomic calcium by polarized incoherent light, and the sunlight-driven photoisomerization of retinal, the first step in animal vision.
Finally, we conclude by summarizing the key results from these studies and highlighting ongoing research directions in this field in section \ref{sec:conc}.

Although we aim for a self-contained exposition with pedagogical introductions to the key tools and concepts used in the study of noise-induced processes, as well as qualitative explanations of phenomena, the reader may find familiarity with a few topics helpful.
Throughout the tutorial, we  use a semiclassical picture of light driven dynamics, where the incident field is treated classically and the materials system is treated quantum mechanically, valid for weak field excitation.  Examples are selected from atomic and molecular systems.
In addition, we  contrast our treatment of noise-induced coherence with the commonly used Pauli rate equations that neglect any quantum phase generated by the noisy field.
A useful outline of these optical theories, as well as their relationship to the quantized fields used in quantum optics, can be found in  Mandel and Wolf \cite{mandel_optical_1995} and Scully and Zubairy \cite{scully_quantum_1997}.
The state and dynamics of these quantum systems will be treated using a density matrix description.
A description of this formalism, including the quantum master equation techniques used to simulate system dynamics, can be found in  Blum \cite{blum_density_2012}, Breuer and Petruccione \cite{breuer_theory_2007}, and May and Kuhn \cite{may_charge_2011}.
Throughout, we  focus  on a pedagogical treatment of the material, omitting some of the technical details of the formalism, simulations and studies.
Where applicable, the interested reader is directed to the original publications for a more complete treatment.

\section{The Density Matrix \& Coherences}
\label{sec:rho}

We  primarily utilize a density matrix description $\hat{\rho}$ of a quantum system to allow for the treatment of stochastically driven processes.
The wavefunction based formalism and its associated Schr\"{o}dinger equation is appropriate only if we possess maximal information on both the system and its driving fields.
This approach only accounts for uncertainty that arises from the quantum nature of the wavefunction $\ket{\psi}$.
If instead the quantum system is subjected to a stochastic driving field, such as noisy incoherent light, its state will also vary stochastically, introducing another source of uncertainty into its observables.
In this case, the exciting field is described by an ensemble, discussed below, which in turn generates an ensemble of wavefunctions.
The resulting statistics of this quantum ensemble can be compactly described by a density matrix
\begin{equation}
  \label{eq:rho}
  \hat{\rho} \equiv\langle \ket{\psi}\bra{\psi}\rangle,
\end{equation}
where $\langle \cdot\rangle$ indicates an average over the ensemble of stochastic exciting fields \cite{liboff_introductory_2003}. The wavefunction $|\psi\rangle$ is of the general form $|\psi\rangle = \Sigma c_i|\phi_i\rangle$ in some basis $|\phi_i\rangle$.  Below, $|e_i\rangle$ denotes eigenstates of the Hamiltonian of energy $e_i$.

Before proceeding, it will be helpful to highlight key aspects of the physical interpretation of density matrices that are of particular interest to this study.
Similar to wavefunctions, the  representation of a density matrix generally depends on the choice of a basis for the Hilbert space $\Hilb$.
Once a basis $\lbrace \ket{n} \rbrace$ is selected, the physical meaning of the diagonal and off-diagonal matrix elements can be readily appreciated.
The diagonal density matrix elements $\rho_{nn}$  describe the probability of observing a quantum system in state $\ket{n}$.
Notably, these values contain no information on the quantum phase between different states.
They only reflect the magnitudes of wavefunction coefficients, i.e.,   the diagonal components of Eq. (\ref{eq:rho}), $\rho_{nn} = \langle |c_n|^2\rangle$.
In particular, if $\hat{A}$ is a system observable with eigenbasis $\lbrace \ket{n} \rbrace$ and eigenvalues $\lbrace a_n\rbrace$, then the probability of measurement outcome $a_n$ is given by the diagonal matrix element $\rho_{nn}$.
Correspondingly, the average value of $\hat{A}$ is
\begin{equation}
  \label{eq:Aave}
  \langle A \rangle = \sum_n a_n \rho_{nn} = \Tr\lbrace \hat{A} \hat{\rho}\rbrace,
\end{equation}
where $\hat\rho$ is expressed in the eigenbasis of $\hat{A}$ in the first equality and $\Tr$ denotes the trace of a matrix, the latter being basis independent.  .

All quantum phase information is stored in the coherences, i.e., the off-diagonal matrix elements $\rho_{ij}$.
It is in these matrix elements that the additional information content of a density matrix approach can be best appreciated.
In the limit where there is no heterogeneity in the wavefunction ensemble (e.g.,  upon excitation by a deterministic field) the system is  in a pure state $\rho = \ket{\psi}\bra{\psi}$.
The coherences are then simply given by the conjugate product $\rho_{ij} = c_i c_j^*=|c_i||c_j|e^{i\phi_{ij}}$ where  $\phi_{ij}$ gives the relative phase between states $\ket{i}$ and $\ket{j}$.

 However, in the general case where, e.g., a stochastic field can generate a heterogeneous ensemble of wavefunctions, i.e., the mixed state case, the magnitude of coherences is no longer purely determined by the state populations $|c_i|$.
Instead, averaging over the complex phase factor can decrease the coherence magnitude.
This situation arises when, e.g., the stochastic field randomizes the relative phase between basis states, termed dephasing.
In the extreme limit where the phase factor $\phi_{ij}$ is uniformly distributed, averaging over this complex number leads to vanishing coherence.
As such, under excitation with stochastic fields,  the magnitude of coherences between two states reflects how well quantum phase information is preserved  and, consequently, how much of a role coherent quantum phenomena such as interference can play in system properties.
The dynamics and steady state values of these coherences is the primary focus in  this tutorial.
In particular, we  consider  situations where a  weak noisy field is able to generate and maintain these coherences.
This regime is of particular interest as it indicates the possibility that quantum effects play a role in system behavior.  Further, it is a natural regime for light-induced biological processes.

One particularly important situation arises in thermodynamic equilibrium.
When a quantum system interacts with a finite temperature bath, the stochastic nature of a noisy bath will eventually generate a thermal ensemble of quantum states.  (Deviations from this behavior are discussed in \cite{pachon_influence_2019}.)
If the system has Hamiltonian $\hat{H}$, then this ensemble is described by the thermal state at temperature $T$
\begin{equation}
  \label{eq:gibbs}
  \hat\rho_{th} = \frac{e^{-\beta \hat{H}}}{\Tr\lbrace e^{-\beta\hat{H}}\rbrace},
\end{equation}
where $\beta = k_B T$, and $k_B$ is the Boltzmann constant.
Expressing Eq. (\ref{eq:gibbs}) in the energy eigenbasis shows that eigenstate $\ket{n}$ has population $\rho_{nn} = e^{-\beta E_n}/\sum_n e^{-\beta E_N}$, consistent with classical thermodynamics.
Moreover,  no coherences remain between non-degenerate energy eigenstates, since $e^{-\beta\hat{H}}$ is diagonal in the eigenbasis of $\hat{H}$.
This indicates that no quantum coherences, and therefore no interference effects, are expected to survive at thermal equilibrium.
Crucially, this result assumes that all fluctuating environment that the system interacts with are at the same temperature $T$.
We will see below that the lack of equilibrium coherence arises from the ``detailed balance'' condition imposed by interaction with a single temperature bath and that the situation becomes far more interesting when the system interacts with systems at different temperatures.

\section{Optical Excitation of Atomic \& Molecular Systems}
\label{sec:excit}

\subsection{Model Hamiltonian \& Dynamics}
\label{sec:ham-dyn}

Consider then the dynamics of quantum systems driven by external fields.
For simplicity, we  take a semiclassical approach where the system of interest is treated fully quantum mechanically while the driving field is treated as a classical force.
That is, the Hamiltonian is given by
\begin{equation}
  \label{eq:Ham}
  \hat{H} = \hat{H}_M - \hat{\bm\mu}\cdot \bm{E}(t)
\end{equation}
where $\hat{H}_M$ is the Hamiltonian of the material (e.g.,  atomic or molecular) system with eigenstates $\ket{\alpha}$ defined by $\hat{H}_M\ket{\alpha} = e_\alpha\ket{\alpha} = \hbar \omega_\alpha\ket{\alpha}$.
The interaction with the electric field $\bm{E}(t)$ is treated in the dipole approximation and is governed by the dipole operator $\hat{\bm\mu}$.
This picture is reliable as well as pedagogically convenient as it transparently shows the influence of the properties of the driving field on molecular excitation and enables an intuitive treatment of the stochastic noisy fields to be  considered.
Moreover, the interaction between natural incoherent radiation (e.g., sunlight or noise) and molecules or atoms are sufficiently weak that this level of theory (including spontaneous emission) is consistent with the fully quantum treatment of the driving field \cite{dodin_light-induced_2019,koyu_steady-state_2020,dodin_coherent_2016,tscherbul_long-lived_2014,kozlov_inducing_2006,agarwal_quantum_2001,agarwal_quantum_1974}.

\begin{figure}[htbp]
  \centering
  \includegraphics[width=0.5\textwidth]{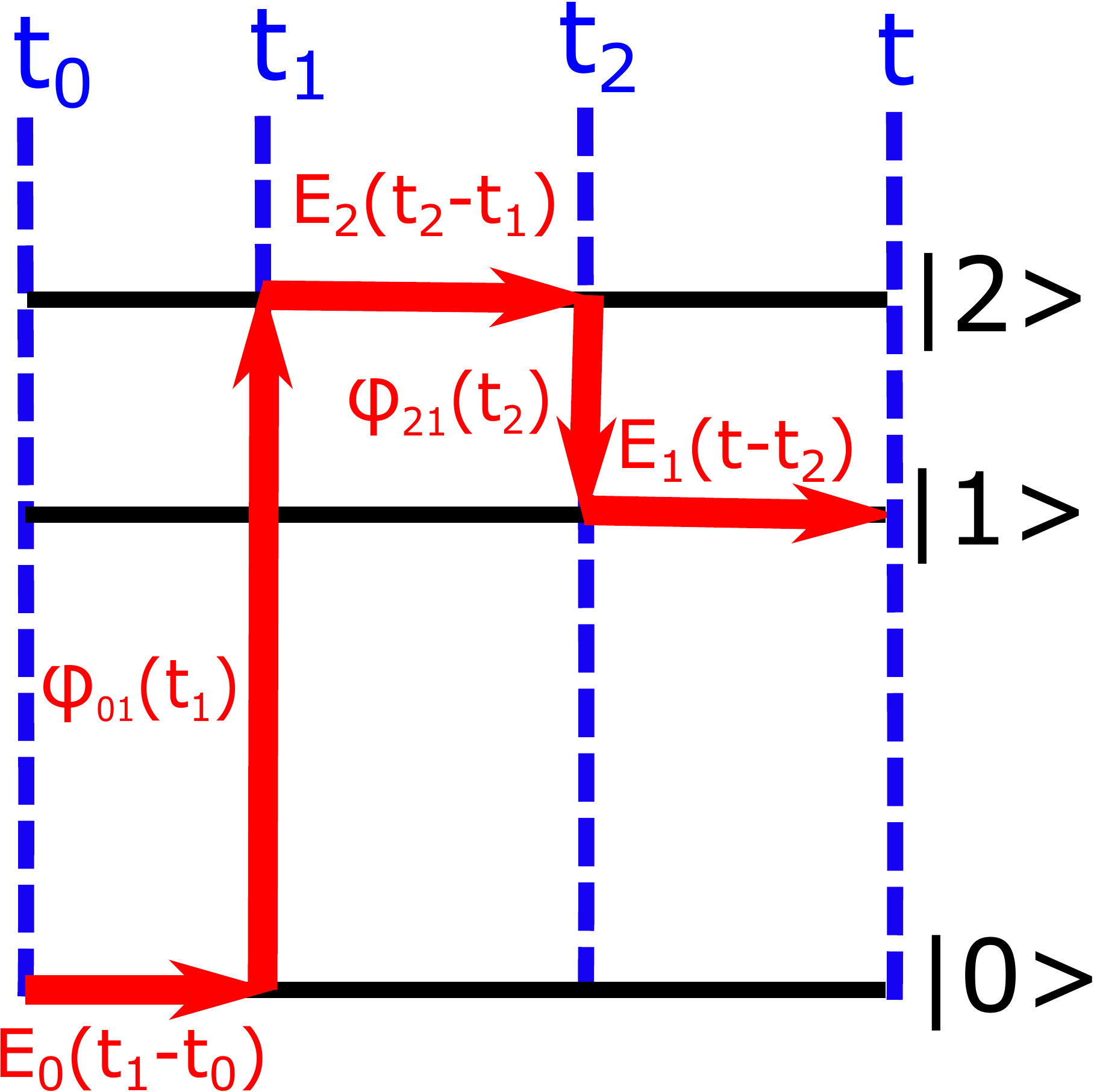}
  \caption{Sample path (red) between energy eigenstates induced by an external field. The phase accrued in each step of the path is shown in red text ($\hbar =1$ units). Horizontal arrows indicate unitary evolution on one eigenstate while vertical lines indicate field induced transitions. The dynamics at final time $t$ are obtained by summing over all such paths.}
  \label{fig:PT-paths}
\end{figure}

The dynamics generated by the electric field can be decomposed as a sum of excitations occurring at different times.
Consider a system with a matter-field Hamiltonian given by Eq. (\ref{eq:Ham}).
In the absence of a driving field $\bm{E}(t)$, the system  undergoes unitary evolution described by the propagator $\hat{U}(t, t_0) = e^{-i\hat{H}_M(t-t_0)/\hbar}$ which accumulates a time dependent phase factor $e^{-i\omega_\alpha(t-t_0)}$ in each eigenstate $\ket{\alpha}$, with no transitions between them.
Transitions between energy levels arise due to interactions with the driving field, and their sequence defines a quantum path through the energy levels of the system.
If the driving field $\bm{E}(t) = |E(t)| e^{i\phi_0(t)}\bm\epsilon$  has an instantaneous phase $\phi_0(t)$ and polarization unit vector $\bm\epsilon$, then it induces a transition between energy eigenstates  $\ket{\alpha}$ and $\ket{\beta}$ with probability $|E(t)|^2|\bm\epsilon \cdot \bra{\alpha}\bm\mu\ket{\beta}|^2$ and with complex phase factor $e^{i\phi_{\alpha\beta}(t)} = e^{i\phi_0(t)}\bm{\epsilon}(t)\cdot \bra{\alpha}\bm{\mu}\ket{\beta}/|\bm{\epsilon}(t)\cdot \bra{\alpha}\bm\mu\ket{\beta}|$. 
Each of these paths carries a complex weight determined by two factors:  (1) the phase factor accumulated through unitary evolution in the time intervals between the field-induced transitions and (2) the phase accrued during the transitions themselves.
One of these paths is sketched schematically in Fig. \ref{fig:PT-paths} alongside the phase accrued in each path segment.
The end state of the system is then obtained by summing over these interfering pathways, accounting for their complex weight.

The same interfering pathways procedures can also be applied to incoherent light provided the fluctuations of the stochastic field are also averaged over.
The resulting dynamics are described by a density matrix averaging over the dynamics induced by the different realizations of the incoherent light, as shown in Eq. (\ref{eq:rho}), with each $\ket{\psi}$ indicating a wavefunction obtained by summing over all interfering pathways.
Generally, we  will be interested in fields with fixed intensity and with a time-dependent fluctuating phase with a correlation time $\tau_c$ referred to as the coherence time.
In this case, the randomization of the field phase will lead to a distribution in the phase accrued during transitions between eigenstates.
Averaging over these different phases will damp out the interference between those excitations separated by times longer than $\tau_c$ since some realizations will show constructive interference while other will show destructive interference, leading to a net loss in interference between pathways.

This time-domain pathways picture is exact provided all paths are enumerated and summed over, and is consistent with a frequency domain analysis.
For example, consider the resonant excitation of a two level system with energy gap $\hbar \omega_0$ excited by a monochromatic laser $E(t) = e^{i\omega t}$.
Consider two different excitation pathways that are identical except for one transition which occurs a time $\Delta t$ later in one of the paths (e.g.,  $t_1 \to t_1 + \Delta t$ in Fig. \ref{fig:PT-paths}).
The earlier excitation will accrue a unitary phase of $\pm \omega_0 \Delta t$ relative to the later excitation.
In addition, the two transitions will acquire different phases during their excitations due to the change in phase of the exciting field, giving a relative excitation phase of $\mp \omega \Delta t$.
If the exciting field is incident on the system for much longer than its period, then the system will experience many oscillations of the field and many of these pathways will be generated.
If $\omega \neq \omega_0$ these pathways will interfere destructively leading to no net transition between the states.
Only if the exciting field is resonant with the transition (i.e. $\omega = \omega_0$) will these pathways always constructively interfere, resulting in transitions between the two states.

These resonance conditions are relaxed if the exciting field is stochastic.
To see this consider a similar two-level system excited by a stochastic field, characterized by a noisy phase $E(t) = e^{i\omega t +\delta \phi_t}$ where $\delta \phi_t$ is a mean zero white noise.
Under these excitation conditions, the phase difference between transitions arising at different times varies randomly with some characteristic correlation time $\tau_c$.
Excitations that are separated by times $\Delta t$ longer than $\tau_c$ will have a random phase relative to one another and therefore cannot interfere.
The loss of destructive interference allows for some transitions to occur away from resonance with $\omega \neq \omega_0$ provided $\omega -\omega_0 < 2\pi/\tau_c$.
Correspondingly, in the frequency domain, this phase noise leads to broadening in the power spectrum of the exciting field with a bandwidth $2\pi/\tau$.

\begin{figure}[htbp]
  \centering
  \includegraphics[width=0.9\textwidth]{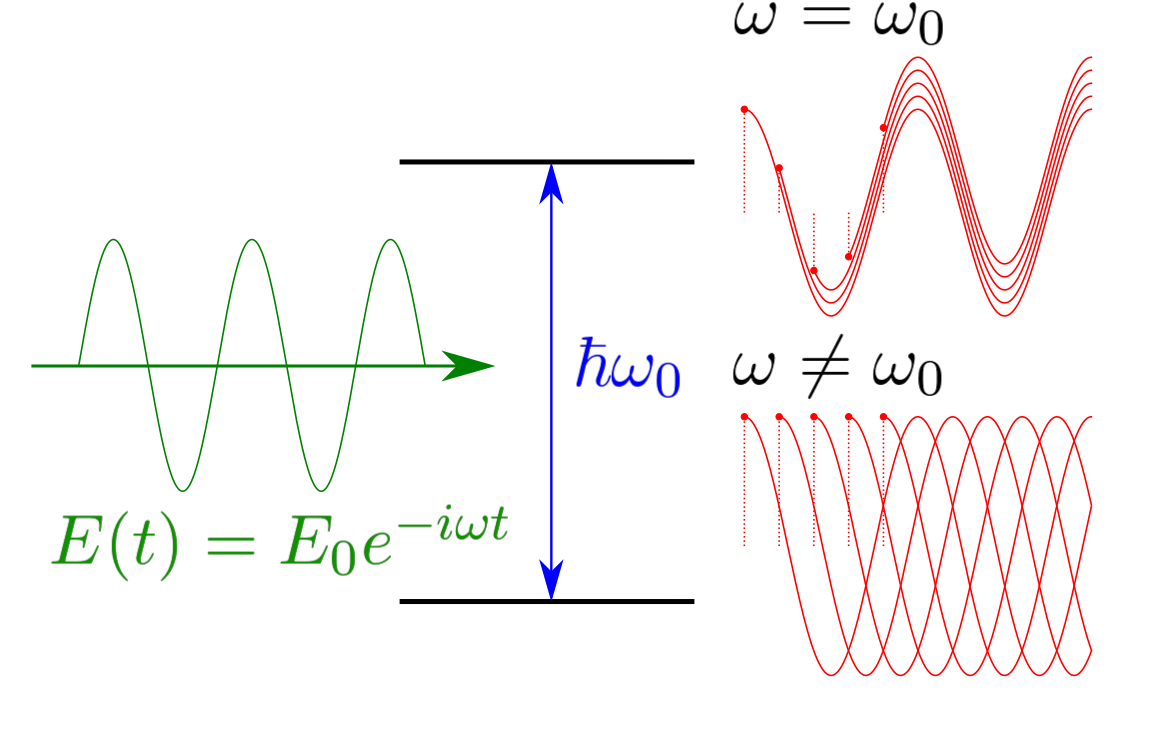}
  \caption{Schematic illustration showing the interference of first order excitation pathways occurring at different times. An exciting field with frequency $\omega$ is shown driving a transition at frequency $\omega_0$. On the left the real part of the excited state coefficient arising from excitations at different times are sketched to illustrate the interference between them. The dashed lines indicate the excitation times drawn. If the field is in resonance with the driving transition ($\omega=\omega_0$, top left) excitations generated at different times will have the same phase, leading to constructive interference and a significant population transfer to the excited states. (Note that excitations at different times are shown with a slight vertical offset for clarity). In contrast, the off-resonant case ($\omega\neq\omega_0$, bottom left), excitations generated at different times will have different phases leading to destructive interference between different first order pathways and no significant population of the excited state.}
  \label{fig:PT1}
\end{figure}

Time-dependent perturbation theory provides a systematic method for estimating field-induced dynamics by enumerating and summing over these excitation pathways.
In the weak coupling limit, the probability of a field driven by $n$ transitions is bounded by $|E_{\mathrm{max}}|^{2n}|\mu_{\mathrm{max}}|^{2n}$ where $|E_{\mathrm{max}}|$ is the maximum field strength and $|\mu_{\mathrm{max}}|$ is the magnitude of the largest transition dipole moment between states.
As such, pathways with more transitions are far less likely in the naturally relevant weak field scenario, and dynamics can be systematically approximated to $n^{th}$ order by considering pathways with $n$ or fewer transitions.
In the scenarios of interest, the exciting field is weak enough that a first order treatment  suffices. The wavefunction then picks up a perturbative correction $\psi^{(1)}(t)$ with

\begin{equation}
  \label{eq:psi_PT}
  \braket{e_n|\psi^{(1)}(t)} = \int^t_{t_0} dt' e^{-i e_n (t'-t_0)/\hbar}\bm{E}(t')\cdot \bra{e_n}\hat{\bm\mu}\ket{g}
\end{equation}
where the light field begins driving a system initially in the ground state $\ket{g}$ at time $t_0$ into the excited eigenstate manifold $\lbrace\ket{e_n}\rbrace$ with energies $\lbrace e_n \rbrace$.
We have set the ground state energy $e_g = 0$.    In this case, in first order, interference arises solely from contributions to the integral in Eq. (\ref{eq:psi_PT}) from different times $t'$.
The interference between first order excitation pathways at different times are sketched in Fig. \ref{fig:PT1} for both the resonant and non-resonant case.
In the resonant case, the phase of the exciting field and therefore the initial phase of the generated excitations, evolves at the same frequency as the phase accrued in the excited state.
Excitations generated at different times are therefore always in-phase with one another leading to constructive interference in the excitation pathways.
When the field is not in resonance with the driven transition, excitations at different times will have different phases leading to destructive interference in the excitation process and very little population of the excited state.

We can take a similar perturbative approach to treat stochastic field dynamics in the density matrix formalism as sums over interfering pathways, provided that we also average over the statistics of the field \cite{chenu_transform-limited-pulse_2016,jiang_creation_1991}.
This yields the corresponding perturbative dynamics after excitation from the ground state $|g\rangle$ (where spontaneous emission and other relaxation mechanisms are neglected here, but included  below):

\begin{equation} \label{eq:rho_tot}
\begin{split}
\rho_{\alpha \beta}(t) \equiv&\: \bra{\alpha}{} \rho(t) \ket{\beta}{} \\
=&\: \frac{ {\bm \mu}_\alpha \cdot {\bm \mu}_\beta^*}{\hbar^2}  e^{-i \omega_{\alpha \beta}t}
\int_{-\infty}^t  d\tau_2 \, e^{i \omega_{\alpha g} \tau_2}   \int_{-\infty}^t  d\tau_1 \,
e^{-i \omega_{\beta g} \tau_1} \\
\times &\langle E(\tau_1)^*E(\tau_2) \rangle \,,
\end{split}
\end{equation}
with $\omega_{\alpha \beta} = (e_\alpha - e_\beta)/\hbar = \omega_\alpha - \omega_\beta$ and ${\bm \mu}_\gamma = \langle
\gamma|\mu|g\rangle$.

Hence, stochastic light-induced processes are described by
$\rho_{\alpha\beta}(t)$ which, in turn, reflects the statistics of the driving field through the first order  correlation function  $\langle E(\tau_1)^*E(\tau_2) \rangle$ of the incident field.
In  quantum optics this is said to be a $g^{(1)}$ process in that it is only sensitive to the degree of first order coherence \cite{mandel_optical_1995}.
\begin{equation}
  \label{eq:corrfunc}
  g^{(1)}(|\tau_1 - \tau_2|)=\langle E^*(\tau_2) E(\tau_1)\rangle/ |E^*(\tau_2) E(\tau_1)|.
\end{equation}
These first order processes do not require a quantum treatment of the light field, and are well described by the semiclassical picture  used  here.
Physically, $g^{(1)}$ describes how long the stochastic field maintains its internal phase relationship, i.e.,  how long the field phase $\phi_0(t)$ defined above remains correlated.
Given the dependence of the pathway phase in Fig. \ref{fig:PT-paths} on this field phase, its correlation time will clearly modify which pathways are able to reliably interfere.  In the relevant case of blackbody radiation, Eq. (\ref{eq:rho_tot}) can be evaluated numerically \cite{chenu_transform-limited-pulse_2016}.  The correlation function  $\langle E(\tau_1)^*E(\tau_2) \rangle$ in Eq. (\ref{eq:rho_tot}) for blackbody radiation is \cite{bourret_coherence_1960} a rapidly falling off function of $(\tau_1  - \tau_2)$ normalized to unity at $(\tau_1  - \tau_2) = 0$.  Below we approximate this correlation function as $A e^{-[\tau_1  - \tau_2]/\tau_c}$     to emphasize both its rapid falloff and its clustering around $(\tau_1  - \tau_2) = 0$. The coherence time $\tau_c$ describes the time over which a stochastic field remains correlated. For natural thermal light   $\langle E(\tau_1)* E(\tau_2)\rangle$ falls to zero within a few fs \cite{bourret_coherence_1960}.

The stochastic phase properties of the incoherent field reduces its ability to reliably generate coherences between energy eigenstates.
This is in contrast to the complex phase of a perfectly coherent light source which is deterministic at all times,
a regime given by $g^{(1)}=1$ in Eqs. (\ref{eq:rho_tot}) and (\ref{eq:corrfunc}), where excitations at all times will reliably interfere.
This type of excitation incorporates interference between transitions occurring at all times $\tau_1$ and $\tau_2$ in Eq. (\ref{eq:rho_tot}) and 
pulsed laser experiments exploit this perfect interference to design pulses that excite specific superpositions at specific times.
By  contrast, the complex phase of incoherent light will randomize over time, stochastically varying the relative phase of transitions occurring at different times.
This finite decay time of $g^{(1)}$ in Eq. (\ref{eq:rho_tot}) will suppress interference between excitations occurring at times separated by more than $\tau_c$,
limiting the possibility of encoding specific superpositions in the spectral and phase properties of the exciting light.
In fact, in the incoherent limit relevant to excitation by natural incoherent sources, $\tau_c \to 0$, preventing excitations occurring at different times from interfering or generating coherences (e.g. natural sunlight has coherence time $\tau_c \sim 1$ fs).
Consequently, the phase of excited superpositions generated by incoherent light cannot be traced back to the exciting field.
However, as we will show below, even in the fully incoherent $\tau_c =0$ limit, noisy fields can still generate excited state coherences, albeit through a very different physical mechanism than that of  coherent light.

\section{Noise-Induced Coherences}
\label{sec:coh}

\subsection{Simultaneous Excitation \& Non-Secular Dynamics}
\label{sec:nic}
In spite of the vanishingly short coherence time of natural incoherent light, these fields are still able to generate coherences and subsequent interference in the energy eigenbasis.
This phenomenon is surprising since it indicates that even when driven by a field with completely uncorrelated phase such as Gaussian white noise, a quantum system can still be  excited into a specific excited superposition  evident in $\rho$ and will subsequently display the dynamical interference characteristic of that superposition.
Nevertheless, this behavior has now been observed in a wide array of theoretical studies including minimal three level models \cite{tscherbul_long-lived_2014,dodin_quantum_2016-1,dodin_coherent_2016}, incoherently driven Calcium atoms \cite{dodin_secular_2018,koyu_steady-state_2020}, quantum heat engines \cite{kozlov_inducing_2006,scully_quantum_2011,dorfman_photosynthetic_2013}, and in simple models for photosynthetic light-harvesting and retinal photoisomerization \cite{tscherbul_excitation_2014,tscherbul_quantum_2015,dodin_light-induced_2019}.
Remarkably, these studies have predicted macroscopic consequences of these noise-induced dynamics including spatial and temporal modulation in the fluorescence emission of atomic Calcium \cite{dodin_secular_2018,koyu_steady-state_2020}, quantum-enhanced heat engine efficiencies that exceed classical limits \cite{scully_quantum_2011} and enhancement in the quantum yield of retinal photoisomerization \cite{tscherbul_quantum_2015,dodin_light-induced_2019}.  They also appear in   atomic  applications as vacuum induced coherences \cite{agarwal_quantum_2012,ficek_quantum_2005}.
Surprisingly these noise-induced coherences can be observed for any  field coherence time, indicating that they do not require the existence of temporal correlations in the exciting field.
To highlight this feature, we  restrict  attention to the fully incoherent limit in this section where the coherence time $\tau_c \to 0$ is much shorter than any other time scale in the system dynamics.

\textit{The superpositions to which we allude,  and resulting quantum interference, appear as coherences in the off-diagonal elements of the density matrix.}
To illustrate the conditions under which noise-induced coherences may appear, consider the off diagonal ($\alpha \neq \beta$) elements of Eq. (\ref{eq:rho_tot}).
First, note that the magnitude of the coherences is scaled by a factor $\bm{\mu}_\alpha\cdot \bm{\mu}^*_\beta$ where $\bm{\mu}_i = \braket{i|\hat{\bm{\mu}}|g}$ is the transition dipole moment from the ground state $|g\rangle$ to excited state $|i\rangle$.
This geometric prefactor describes the alignment of the transition dipole moments of the two transitions and quantifies the probability that a randomly chosen electric field polarization is able to excite both the $\ket{g}\to \ket{\alpha}$ and the $\ket{g}\to\ket{\beta}$ transition.
For example, if the transition dipole moments are colinear, then any polarization is equally likely to excite either transition.
In contrast, if they are orthogonal, then the more likely a polarization is to excite one transition (i.e. the better the electric field is aligned with the transition dipole moment) the less likely it is to excite the other since it is closer to being orthogonal to its transition dipole moment.

Notwithstanding these geometric considerations, the question remains how an electric field with an instantly randomizing phase is able to preferentially excite a given superposition in the excited state.
To address this question, consider the time integral in Eq. (\ref{eq:rho_tot}).
For both population $\alpha = \beta$ and coherence $\alpha \neq \beta$ density matrix elements, the interference between transitions occurring at times $\tau_1$ and $\tau_2$ is attenuated by the $g^{(1)}$ correlation function, indicating that only excitation times separated by less than the coherence time of the field can interfere.
This indicates that in the incoherent limit where $\tau_c \to 0$, only simultaneous excitations with $\tau_1 \approx \tau_2$ contribute on average to the system dynamics.
For non-simultaneous excitations, the phase of the exciting field has fully randomized leading to a full cancellation of in-phase and out-of-phase excitations, where the phase of the excitation is determine by the now randomized phase of the exciting field.
However, for  simultaneous excitations, the field phase is the same for both transitions since they are driven by the same field at the same time and therefore with the same phase.
In this case, the field can introduce no relative phase between simultaneous transitions and therefore the only phase that can be introduced appears in the geometric prefactor $\bm{\epsilon}(t)\cdot \bra{i}\bm{\mu}\ket{j}/|\bm{\epsilon}(t)\cdot \bra{i}\bm\mu\ket{j}|$ which reflects the orientation of the molecular system with respect to the electric field polarization.
The net effect of this relative phase is contained in the geometric factor  $\bm{\mu}_\alpha \cdot \bm{\mu}^*_\beta$ discussed above.

\textit{In summary, noisy fields can still induce coherences due to interference between simultaneous excitations at a time $t' = \tau_1=\tau_2$, from one ground state to multiple different excited states, and the phase of this coherence is determined by the relative orientation of the transition dipole moments of the excited states.}
These simultaneous excitations are  being continuously generated at all times $t' \in  [-{\infty}, t]$ and the dynamics at time $t$ are given by the interference of simultaneous excitations occurring at all of these different times $t'$,  since the double integral in Eq. (\ref{eq:rho_tot}) over excitations reduces to a single integral over simultaneous excitations occurring at time $t'= \tau_1=\tau_2$. 
To see this, consider a coherence function of the form $g^{(1)}(t) = \exp(-t/\tau_c)$ .
Substituting this expression into Eq. (\ref{eq:rho_tot}) using $\langle E^*(\tau_1)E(\tau_2)\rangle = |E^*(\tau_2)E(\tau_1)|g^{(1)}(|\tau_1-\tau_2|)$, the correlation function will exponentially damp virtual excitation times that differ by more than the coherence time $\tau_c$.
In the case of sunlight, $\tau_c\sim 1 \mathrm{fs}$ is much faster than the system timescales, allowing us to approximate $g^{(1)}(|\tau_1-\tau_2|) \approx \delta(\tau_1-\tau_2)$, reducing Eq. (\ref{eq:rho_tot}) to a single time integral
\begin{equation}
  \label{eq:oned}
  \rho_{\alpha\beta}(t) = \frac{{\bm \mu}_\alpha \cdot {\bm \mu}_\beta^*}{\hbar^2}  e^{-i \omega_{\alpha \beta}t}
\int_{-\infty}^t  dt'  e^{i \omega_{\alpha \beta} t'} |E(t')|^2
\end{equation}
This physical mechanism, then, is entirely different from the coherences generated by coherent excitation.

One important consequence of this noise-driven mechanism is that simultaneous excitations occurring at different times $t'$ will all have the same static phase determined by ${\bm \mu}_\alpha \cdot {\bm \mu}^*_\beta$, which cannot compensate for the phase accrued during unitary evolution.
This behavior is quite different from resonant excitation by coherent light, where the unitary phase is exactly compensated for by the evolving phase of the electric field. This behavior plays a significant role in determining the regimes where appreciable noise-induced coherence can be generated.
To see this, consider the complex exponentials in Eq. (\ref{eq:oned}) which describe the unitary evolution of the system after excitation from the ground state.
Noise-induced coherences are generated by simultaneous excitations from the ground state to two different excited eigenstates $\alpha \neq \beta$ at time $t'$.
During the transition, these excitations interact with the same field and therefore pick-up the same phase in the transition contained in the phase of ${\bm \mu}_\alpha \cdot {\bm \mu}^*_\beta$.
When the two states are non-degenerate, they will accrue a relative phase of $e^{-i\omega_{\alpha\beta}(t-t')}$ due to unitary evolution in the excited state manifold.
In the absence of relaxation processes such as dephasing and population decay to the ground state, simultaneous excitations arising at different times $t'$ will eventually destructively interfere, leading to the loss of noise-induced coherence at times longer than $1/\omega_{\alpha\beta}$ and leading to the complete loss of noise-induced coherence at long times.
If we are only interested in dynamics on time-scales longer than these energy splittings in the absence of additional environmental relaxation \footnote{The dephasing due to the stochastic fluctuation of field phase captured by the decaying $g^{(1)}$ is explicitly treated in this approach.},  the noise-induced coherences can be neglected, allowing for the decoupling of the dynamics of the populations and of the coherences, with the populations then governed by the Pauli rate laws, a result termed the secular approximation.

However, if relaxation processes contribute (e.g., relaxation due to a surrounding environment) the situation changes.  Relaxation processes that, e.g.,  either lead to the decay of the excitation to the ground state or to randomize the phase of excited state superpositions through energy level fluctuations can remove excitations from this ensemble before they destructively interfere, allowing coherence to build up in the excited state manifold.
If these relaxation processes occur on a timescale $\tau_R$ then an excitation generated at time $t'$ can only contribute to interference until a time $t' + \tau_R$ after which it can no longer contribute since it will have either decayed to the ground state or accrued some additional random phase due to unitary evolution on a fluctuating energy level.
The consequent loss of interference can then prevent the complete destructive interference of coherence that would occur in the absence of relaxation.
In particular, if the states are nearly degenerate relative to the relaxation time scale, i.e.,   $\omega_{\alpha\beta}\tau_R \ll 1$,  the noise-induced coherences can contribute substantially to subsequent evolution of the population (e.g., through interference in system relaxation processes). Coherences and populations can then no longer be decoupled, preventing a role for the secular approximation.
This non-secular regime commonly appears in many atomic and molecular systems.
For example, large molecules with $\gtrsim 10$'s of atoms typically have dense manifolds of closely spaced excited vibronic states that fall within the non-secular regime \cite{grinev_realistic_2015,tscherbul_long-lived_2014,tscherbul_quantum_2015,dodin_light-induced_2019,dodin_coherent_2016,dodin_quantum_2016-1}.
Similarly, the angular momentum $p$ states of atomic systems in weak magnetic fields, discussed below, are typically nearly degenerate and also fall within this regime \cite{koyu_steady-state_2020,dodin_secular_2018}.

\subsection{Steady State Processes}
\label{sec:steady}

In addition to their part in enabling non-secular dynamics, relaxation processes also play an important role in determining the steady state that encodes the physics of incoherent excitation.
These relaxation process compete with the incoherent excitation described above, and with unitary dynamics, to determine the steady state.
For simplicity, we  restrict  attention  to population relaxation processes through which the system decays from each excited state $\ket{\alpha}$ to the ground state $\ket{g}$ at a rate $\gamma_\alpha$.
This rate captures the effect of both radiative (i.e.,  spontaneous emission) and non-radiative (e.g., induced by the phonon environment) decay.

To make the discussion more concrete, it is helpful to consider the simplest system that is able to support noise-induced coherence.
Such a system requires a single ground state $\ket{g}$ that can be driven into two excited states $\ket{e_1}$ and $\ket{e_2}$.
The energy difference between the ground state and $\ket{e_1}$ is given by $\hbar \omega_0$, where $\omega_0$ is typically in the UV or visible range in molecular and atomic systems, while the two excited states differ in energy by $\Delta \ll \hbar\omega_0$.
The system is continuously pumped from the ground state into the  excited states $\ket{e_i}$ by incoherent light at a rate $r_i = (\hbar \mu_{e_i}^2\omega_0^3 \overline{n}(\omega_0))/(3\pi \epsilon_0 c^3)$ where $\mu_{e_i}$ is the transition dipole moment from the ground state to $\ket{e_i}$, $\epsilon_0$ is the permitivity of free space and $c$ is the speed of light.
The constant $\overline{n}(\omega)$ measures the intensity of the incoherent light at frequency $\omega$ and corresponds to the mean number of photons at that frequency.
If the incoherent light originates from a blackbody source at temperature $T$, then $\overline{n}(\omega) = \left(e^{\hbar\omega/(k_BT)}-1\right)^{-1}$, where $k_B$ is Boltzmann's constant.

\begin{figure}[ht!]
  \includegraphics[width=0.5\textwidth]{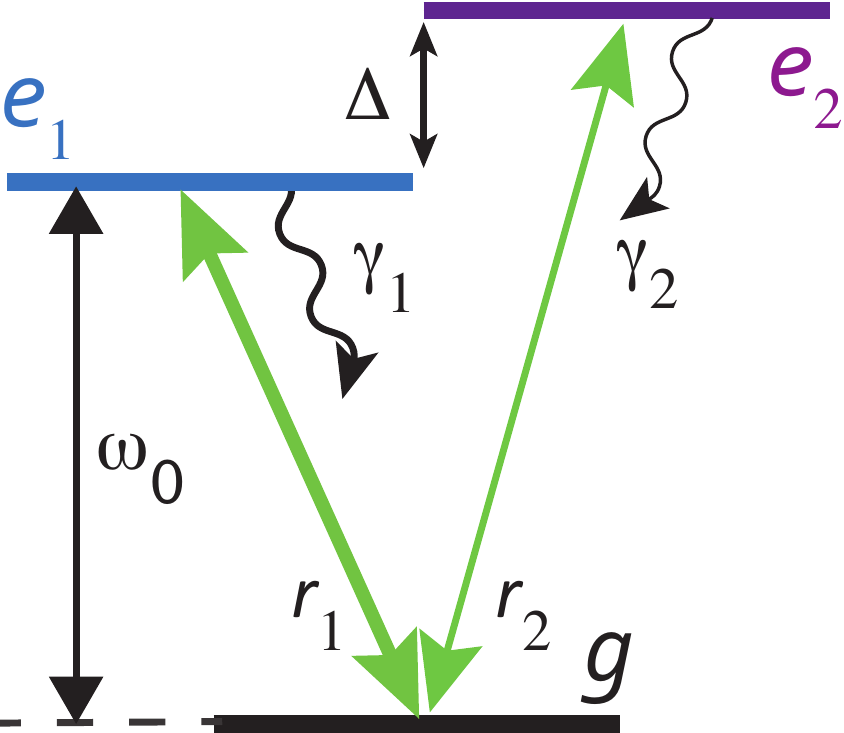}
  \caption{Schematic representation of a V-type system.
    $\Delta$ is the excited state splitting, $\gamma_i$ is the excited state decay rate and $r_i$
    is the incoherent pumping rate of excited state $\ket{e_i}$.  From Ref. \cite{dodin_quantum_2016-1}}
  \label{fig:schem}
\end{figure}

This $V$-system model is sketched schematically in Fig. \ref{fig:schem} and its dynamics are described by the following Quantum Master Equations\cite{tscherbul_long-lived_2014,tscherbul_partial_2015,dodin_coherent_2016,dodin_quantum_2016-1,dodin_secular_2018}: 
\begin{subequations}
  \label{eqs:psbrme}
  \begin{equation}
    \label{eq:rhoeiei}
    \dot{\rho}_{e_ie_i} = r_i \rho_{gg} - (r_i + \gamma_i)\rho_{e_ie_i} - p (\sqrt{r_1r_2} +\sqrt{\gamma_1 \gamma_2}) \rho_{e_1e_2}^R
  \end{equation}
  \begin{align}
    \label{eq:rhoe1e2}
    \dot{\rho}_{e_1e_2} =& p\sqrt{r_1r_2}\rho_{gg} -i\Delta \rho_{e_1e_2} -\half(r_1 + r_2 + \gamma_1 + \gamma_2)\rho_{e_1e_2}\\
                          &- \frac{p}{2}(\sqrt{r_1r_2}+\sqrt{\gamma_1\gamma_2})( \rho_{e_1e_1} + \rho_{e_2e_2})
  \end{align}
\end{subequations}
where $p =\bm{\mu_1}\cdot\bm{\mu_2}/|\mu_1||\mu_2|$ is a normalized alignment parameter of the dipole operators,  $\rho_{e_1e_2}^R $ denotes the real part of the excited state coherence $ \rho_{e_1e_2}$, and the population of state $|k\rangle$ is $\rho_{kk}$.
The separation between the ground and excited state $\hbar\omega_0$ is assumed to be large enough to satisfy the secular approximation, decoupling the ground-to-excited state coherences $\rho_{ge_i}$ from the remaining density matrix elements.
These equations of motion were derived using \cite{tscherbul_partial_2015} the Partial Secular Bloch Redfield (PSBR) equation,  and the dynamics of this model have been thoroughly characterized \cite{tscherbul_long-lived_2014,dodin_quantum_2016-1,dodin_coherent_2016}.
Note that in the absence of non-radiative relaxation processes all population relaxation is driven by spontaneous emission at a rate $\gamma_i$ and stimulated emission at rate  $r_i$.
These rates are related by $r_i = \gamma_i \overline{n}(\omega_0)$.

Consider the physical meaning of each term in Eq. (\ref{eqs:psbrme}).
The terms proportional to the ground state population $\rho_{gg}$ describe the incoherent excitation from the ground state into a  mixture of the  excited states $\ket{e_1}$ and $\ket{e_2}$ and into their superposition $\ket{e_1} \pm \ket{e_2}$ where the sign and magnitude of the superposition is determined by  $p$.
That is, the probability of exciting into the superposition rather than into an incoherent mixture of the excited states with no relative phase is determined by the alignment parameter $p$, as discussed above.
For well aligned excitations $p = \pm1$  the system is excited entirely into the superposition of excited states.
Systems in the excited state will then undergo unitary dynamics that propagate the phase of the coherence $\rho_{e_1e_2}$ through the complex term $-i\Delta\rho_{e_1e_2}$ in Eq. (\ref{eq:rhoe1e2}).
In addition, excitations will relax into the ground state, i.e.,  the populations will decay due to spontaneous and stimulated decay processes through the incoherent decay term $-(r_i +\gamma_i)\rho_{e_ie_i}$ in Eq. (\ref{eq:rhoeiei}).
Similarly, the coherences will also decay as excited state superpositions relax incoherently to the ground state at a rate given by the average decay rate $(r_1 +r_2 +\gamma_1 + \gamma_2)/2$.
The remaining contributions in Eq. (\ref{eqs:psbrme}) that couple the coherences and population describe the interference effects in the population relaxation pathways.

\begin{figure}[ht]
	\includegraphics[width=\textwidth, trim = 0 0 0 0]{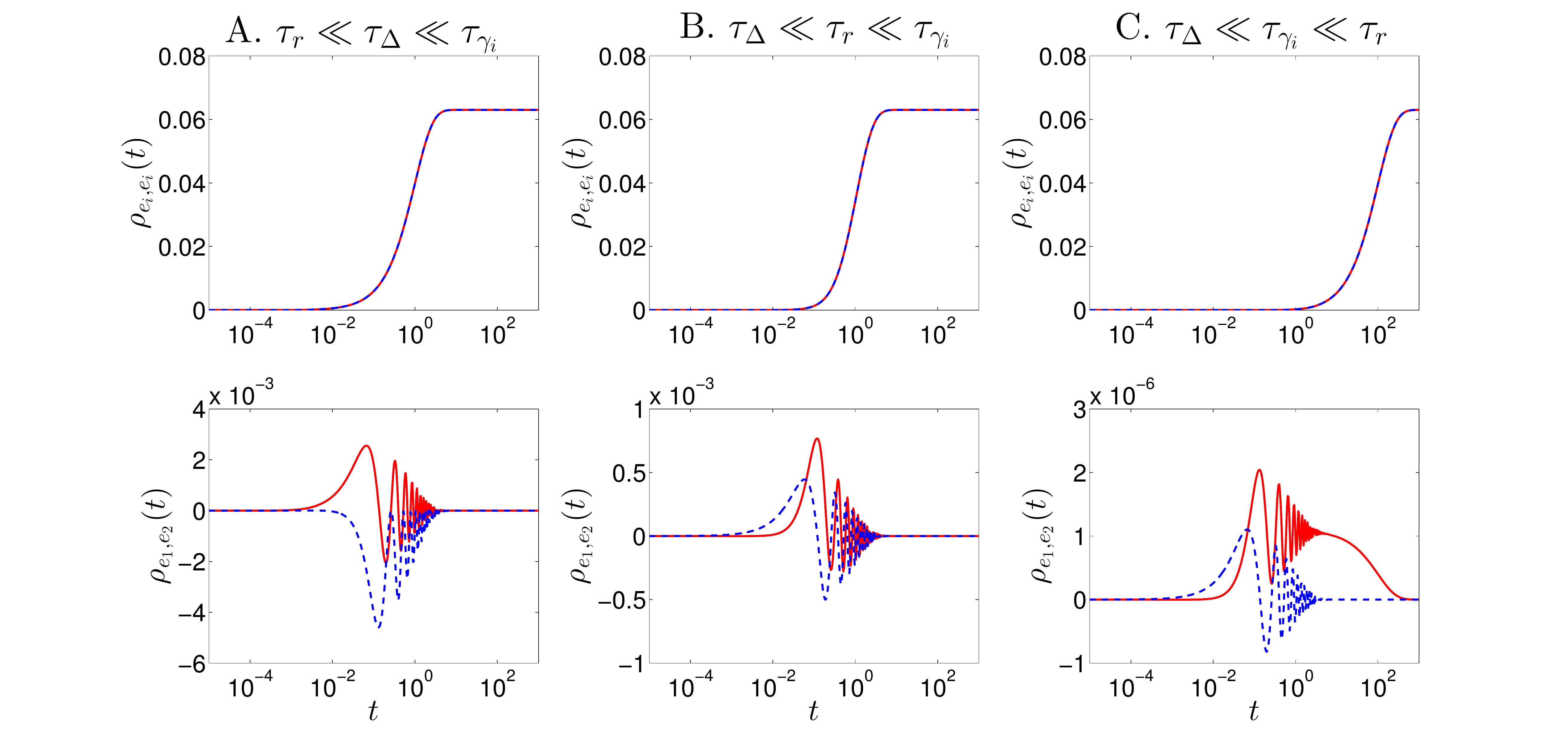}
	\renewcommand{\figurename}{Fig.}
	\caption{Evolution of populations  (upper panels) and coherences (lower panels) of an
``underdamped"
	V-system (i.e.,  where $\Delta/\gamma \gg 1$). Here
	$\gamma_1=1.0=\gamma_2=\gamma$ and $\Delta=24.0$. Three different
	turn-on regimes are shown here. Panels A show the ultrafast turn-on of the field with
$\tau_r=0.024\tau_\Delta$ while Panels B and C show the intermediate ($\tau_r = 24\tau_\Delta$)
and slow ($\tau_r =100\tau_\gamma$) turn-on regimes respectively. Note the difference in
ordinate scales for the coherence plots. Solid red lines indicate the real part of the
coherence $\rho_{e_1e_2}^R$ with the imaginary part $\rho_{e_1e_2}^I$ indicated by the dashed
blue line.  From Ref. 48.}
\label{amrslowfig:2}
\end{figure}

\begin{figure}[ht]
	\includegraphics[width=\textwidth, trim = 0 0 0 0]{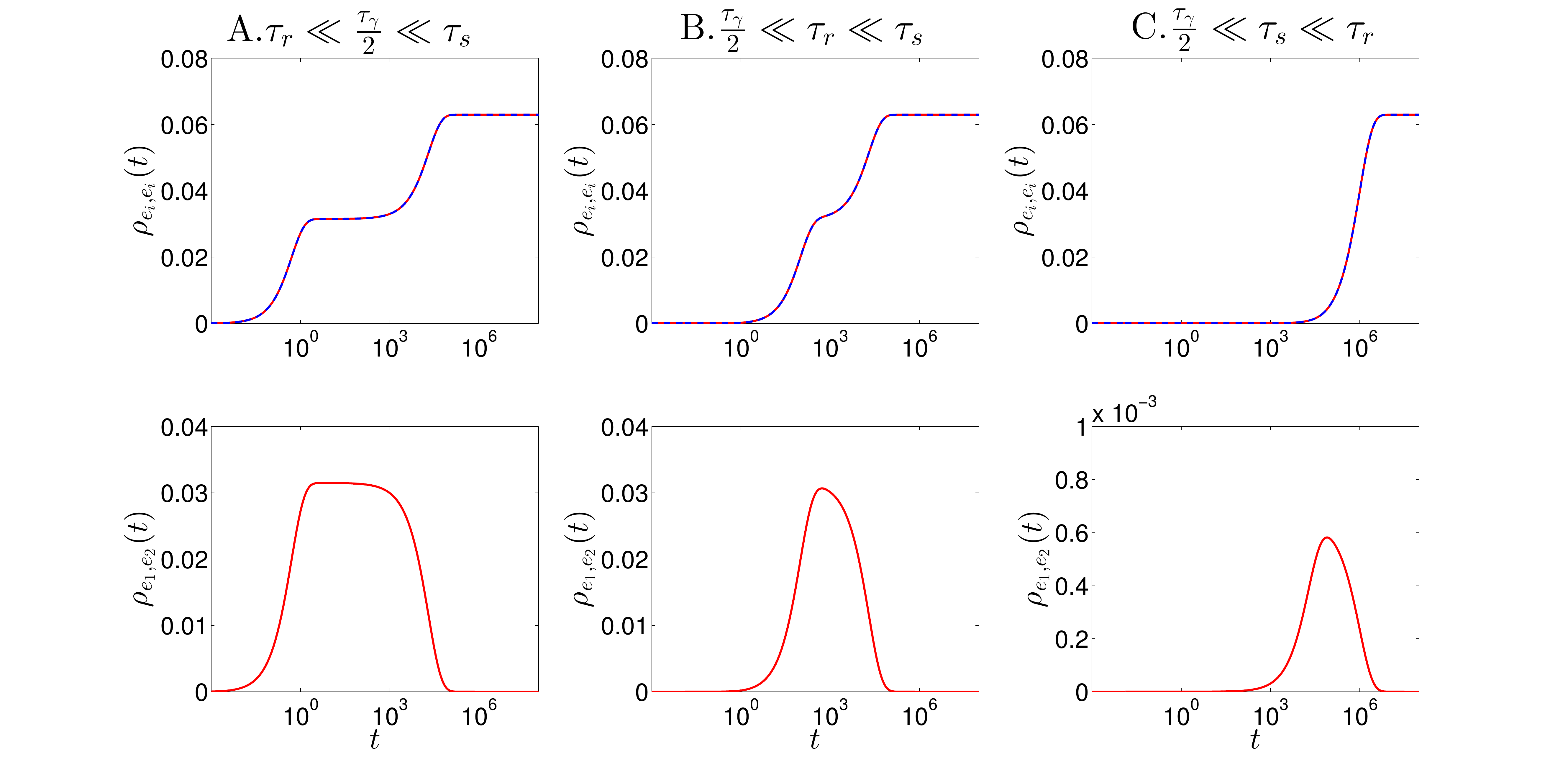}
	\renewcommand{\figurename}{Fig.}
	\caption{Evolution of populations  (upper panels) and coherences (lower panels)  of a
V-system in the
	``overdamped" region (i.e., where $\Delta/\gamma \ll 1$).
	Here $\gamma_1=1.0=\gamma_2$ and $\Delta=0.001$. Three different turn-on regimes are
shown. Panels A  show the ultrafast turn-on of the field with $\tau_r=10^{-3}\tau_\gamma$ while
Panels B and C show the intermediate ($\tau_r=100\tau_\gamma=5\times10^{-5}\tau_s$) and slow
($\tau_r= 20\tau_s$) turn-on regimes, respectively. Note the difference in ordinate scales for
the different coherence plots.  From Ref. 48.}
\label{amrslowfig:3}
\end{figure}

The master equations in Eq. (\ref{eqs:psbrme}) display two distinct dynamical regimes.
For simplicity, in this section we will consider the dynamics in the absence of non-radiative decay pathways, giving the relationship $r_i = \gamma_i \overline{n}(\omega)$.
The first ``underdamped'' regime corresponds to the situation where $\Delta/\gamma > 1$.
In this limit, the excitations generated by the incoherent driving field remain in the excited state manifold for long enough to undergo significant unitary evolution.
The resulting coherences will then oscillate several times before decaying as excitations generated at different times interfere destructively as described by the secular regime in Section \ref{sec:nic}.
Correspondingly, we see that the population dynamics decouple from the coherences and simply show an exponential rise towards a steady state as predicted by incoherent Pauli rate laws.
These dynamics are illustrated for a system where the exciting field is abruptly turned on at time $t_0=0$ in the left panels of Fig. \ref{amrslowfig:2}. (Note $\tau_r$ is the turn-on time of the radiation, $\tau_\Delta = \hbar/\Delta$ and $\tau_{\gamma_i} = \hbar/\gamma_i$).  In this underdamped regime, the dynamics of the population are well-described by a Pauli rate-law equation, while the coherences decrease with increasing excited state splitting $\Delta$. This shows that the partial secular equations of motion we consider naturally approach the secular dynamics in the appropriate large $\Delta$ limit.
However, although the coherences become quite small with increasing $\Delta$ they never fully disappear and may modify the spatial profile of emitted light as we will discuss below \cite{dodin_secular_2018}.

By contrast, the ``overdamped" regime occurs well in the non-secular regime where $\Delta/\gamma < 1$.
In this case, the superpositions generated by the incoherent light have unitary dynamics that are much slower than the excited state lifetime.
The resulting dynamics are then quite different from the ``underdamped'' regime, with the incoherently excited superposition continuously being repopulated and decaying before unitary dynamics is able to substantially modify the coherent phase.
As a result, the destructive interference between excitations occurring at different times is significantly suppressed since nearly all excitations will remain very close to the initially excited superposition until they decay.
Consequently, the system shows an exponential rise into the initially excited superposition, which then survives for a substantial amount of time before being replaced by the incoherent mixture of excited states predicted by Pauli rate laws at long times, as shown in the left panels of Fig. \ref{amrslowfig:3}.
The eventual decay in the coherences can be attributed to the slow build up of excitations that happen to survive for much longer than the excited state lifetime and therefore build up sufficient unitary phase to destructively interfere with the newly generated excitations.
Of course, the survival of an excitation for long enough to accrue enough unitary phase to destructively interfere is quite rare and therefore the build up of these destructively interfering pathways and consequent loss of coherence is slow.
However, if the excitation continues long enough, even these rare events will eventually accumulate.
We also note that since the same superposition is continuously being excited, the coherences are much larger than in the underdamped regime and, in fact are comparable to the populations, indicating that the quantum system is in fact in the incoherently excited superposition.

The dynamics that we have discussed thus far have assumed that the light is suddenly turned-on at some initial time $t_0 =0$.
Clearly, this assumption is quite different from natural conditions where the intensity of light will typically increase extremely slowly on molecular and atomic time-scales.
For example, even the ``blink of an eye'' which  defines a characteristic time scale for light intensity in the human eye typically occurs on the millisecond timescale, about 10 orders of magnitude slower than the timescales of molecular dynamics.
Moving to such systems with more realistic turn-on timescales will significantly change the expected dynamics.
To see this, consider the interfering pathways description for the two dynamical regimes discussed above.
A central consideration in this description is how the phase of an excitation at a given time relates to the phases of excitations generated at earlier times.
The sudden turn-on of the incoherent light discussed above imposes a stark temporal asymmetry in the generation of these superpositions where excitations generated at very early times near $t=0$ have no earlier excitations with which to destructively interfere.
In contrast, if the intensity of the exciting light is very slowly increased starting at $t=0$ then the extremely low intensity at early times will lead to the generation of very few excitations that display this temporal asymmetry.
If the intensity increases slowly then any excitation generated at a time $t >0$ will interfere with excitations generated at earlier times.
Since the change in intensity is very slow, there will have been nearly as many excitations generated shortly before $t$ as there were at time $t$ indicating that at all times the dynamics will reflect the interference with excitations generated at earlier times.
This implies that if the turn-on is sufficiently slow, then the destructive interference with excitations at earlier times will entirely suppress the transient coherences predicted in both the over-damped and under-damped regime. This is seen in the center and right panels of Figs. \ref{amrslowfig:2} and \ref{amrslowfig:3}.
Consequently, under realistic excitation conditions, the only coherences that will be observed in incoherently driven atomic and molecular systems will be those that survive in the steady state.  (This conclusion is rigorously proven in Ref. \cite{dodin_generalized_2021} via a generalized adiabatic theorem.)

\section{Transport and Steady State Coherences}
\label{sec:transport}

\subsection{Detailed Balance \& Thermal Equilibrium}
\label{sec:eqm}

In section \ref{sec:steady} we considered the dynamics of an incoherently driven system in the absence of a non-radiative environment.
In this case, all transitions between system eigenstates are driven by interactions with a single bath with one characteristic temperature.
Correspondingly, at long times, the system tends to  approach a thermal equilibrium state, as described by Eq. (\ref{eq:gibbs}).
Of particular note, this equilibrium Gibbs state has no coherences between system eigenstates.
In combination with the observation that realistic turn-on times suppress transient coherences, this indicates that a single incoherent bath is unlikely to generate significant coherence between the energy levels of molecular and atomic systems under natural conditions.  (For deviations from this expectation, see Ref \cite{pachon_influence_2019}).

In many common light-driven processes, however, a quantum system typically interacts with multiple baths at different temperatures.
For example, a molecular system may be excited by sunlight, a hot bath at temperature $T_H \sim 5800$ K, and dissipate its energy into a relatively cold condensed phase environment, e.g., at $T_C = 278$ K, while an atomic system may be excited by a polarized incoherent beam and dissipate its energy through isotropic spontaneous emission near $0$ K.
In these scenarios, the quantum system will eventually reach a non-equilibrium steady state characterized by a net flux of energy from the hot bath to the cold bath through the system.
As we  show below, the resulting departure from equilibrium can have a profound effect on system coherences, revealing an important connection between coherences in the non-equilibrium steady-state and energy transport through the system.

Consider first what constraints are implied when all system transitions are driven by a single bath and how these ensure that the dynamics of the system will eventually decay to an equilibrium state.
To characterize these constraints consider the transition rates of incoherent excitation and decay in Eq. (\ref{eqs:psbrme}).
If all transitions are driven only by the incoherent  light,  and non-radiative processes do not play a role, then the excitation and decay rates are related by the detailed balance condition $r_i = \gamma_i \overline{n}(\omega)$.
This condition originates \cite{scully_quantum_1997} from the secular Pauli rate law equations and defines the constraints that the population transfer rates must satisfy to ensure that they properly approach equilibrium populations.
Notably, these constraints were obtained by explicitly excluding coherences and therefore there is no reason to expect that they will impose a constraint on coherence dynamics in the non-secular regime, and certainly not that they will guarantee their eventual decay.

In order to determine what parameters determine the steady state coherences in an incoherently driven system it is useful to return to the simple V-System model in Eq. (\ref{eqs:psbrme}) and consider its steady-state.
In particular, we allow the excitation $r_i$ and spontaneous decay $\Gamma_i$ rates to vary independently and consider their implication on the steady-state coherence.
We have slightly modified the notation for the decay rates using capital $\Gamma_i$ to emphasize that the decay rate is not necessarily related to the excitation rate via detailed balance.
While it is possible to obtain a closed form analytical solution for the steady state in the general case of arbitrary parameters, the resulting expression is extremely unwieldy and difficult to parse.
We can simplify the expression substantially while retaining the essential physics by considering the case where $p=1$ and $r_1 = r = r_2$.
This simplification gives the steady state coherences \cite{dodin_light-induced_2019}
\begin{subequations}
  \label{eq:steadyState}
  \begin{equation}
    \label{eq:SSRe}
    \lim_{t\to\infty}\rho_{e_1,e_2}^R(t)=\frac{\sqrt{\Gamma_1\Gamma_2}}{\Gamma_1+\Gamma_2}\left(\frac{(\sqrt{\Gamma_1}-\sqrt{\Gamma_2})^2}{(\sqrt{\Gamma_1}-\sqrt{\Gamma_2})^2+2\Delta}\right)
  \end{equation}
  \begin{equation}
    \label{eq:SSIm}
    \lim_{t\to\infty}\rho_{e_1,e_2}^I(t)=-\frac{\Delta}{r+\half(\Gamma_1+\Gamma_2)}\lim_{t\to\infty}\rho_{1,2}^R(t).
  \end{equation}
\end{subequations}
In this simplified case, we see that the system can show non-zero coherences when $\Gamma_1 \neq\Gamma_2$,  whereas coherences vanish when the decay rates are equal.
In general, this condition can be generalized to arbitrary excitation rates and alignments to show that coherences can be generated only if $p\neq 0$ and $r_1/\Gamma_1 \neq r_2/\Gamma_2$.
However, if the excitation and decay rates are related by a detailed balance condition then equality of these ratios is guaranteed.
Therefore, detailed balance in excitation and decay rates, i.e., equilibrium  does in fact guarantee, in this pedagogical example, the eventual decay of coherences between energy eigenstates.

\subsection{Transport Induced Coherences}
\label{sec:tic}

The difference in excitation and decay rate ratios required to establish coherences in the steady state indicates the presence of energy transport through the system.
To see the connection between transition rates and energy transport, note that if the ratios are not equal, say $r_1/\Gamma_1 > r_2/\Gamma_2$, then overall more energy enters the quantum system through the $\ket{g} \leftrightarrow \ket{e_1}$ transition and more energy leaves through the $\ket{g} \leftrightarrow \ket{e_2}$ transition.
This imbalance in energy flux through the two eigenstate transitions is indicative of a net transport of energy through the system.
Unfortunately, energy transport does not appear transparently in the eigenstate basis since it is often subtly encoded in the details of interference between different transition pathways.

\begin{figure}[ht!]
  \includegraphics[width=0.9\textwidth]{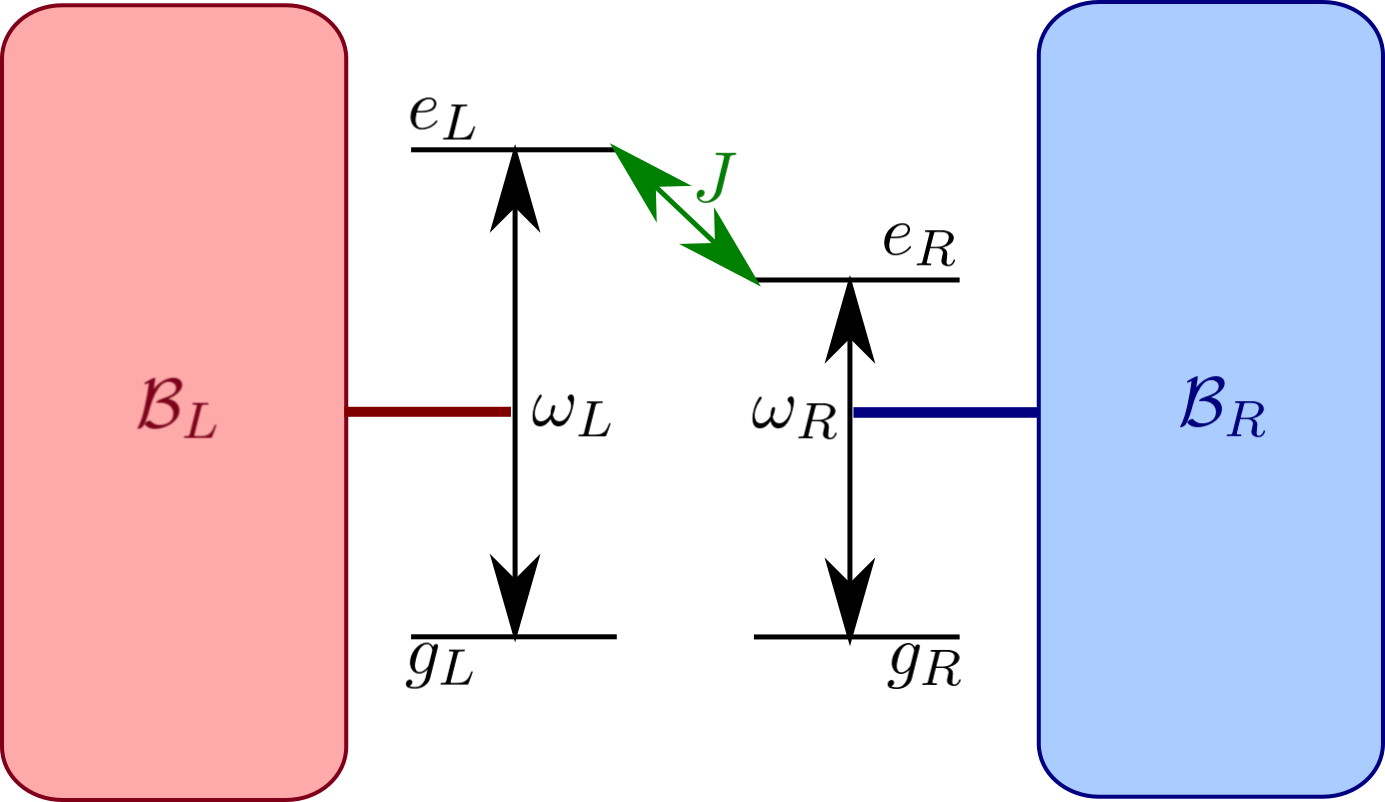}
  \caption{Schematic representation of a two qubit dimer model. The left qubit with transition frequency $\omega_L$ is coupled to one bath $\mathcal{B}_L$ while the right qubit with transition frequency $\omega_R$ is coupled to an independent bath $\mathcal{B}_R$. The two qubits are coupled with a constant $J$ allowing for energy flow between them.}
  \label{fig:dimer}
\end{figure}

In order to clarify the relationship between energy transport and steady state coherence, it is helpful to move to a representation where the flux of energy into and out of the system appears more directly.
To accomplish this, we  consider a standard dimer model for transport between two heat baths and then establish its equivalence to the $V$-system model considered above \cite{koyu_steady-state_2020}.
The dimer model is constructed as follows.
Consider a pair of two-level systems (i.e. qubits), labeled $\mathcal{L}$ and $\mathcal{R}$ to denote the left and right qubit respectively.
Each qubit has a ground to excited state energy gap $\hbar\omega_{\mathcal{L}}$ and $\hbar\omega_{\mathcal{R}}$ and is linearly coupled to its own bath $\mathcal{B}_{\mathcal{L}}$ and $\mathcal{B}_{\mathcal{R}}$ that may be at different temperatures.
If they are at different temperatures, we will use the convention that the left bath is at a higher temperature such that energy flows overall from left to right.
The interaction of each qubit with its bath can lead to transitions from the ground to excited state as energy flows into and out of the qubit from the bath.
In addition, the two qubits are coupled by a constant $J$ that allows for the flow of energy between them.
This model is illustrated schematically in Fig. \ref{fig:dimer}

The system Hamiltonian for this model can then be expressed as
\begin{equation}
  \label{eq:H2qub}
  \hat{H}_M = \hbar\omega_{\mathcal{L}}\ket{e_{\mathcal{L}}g_{\mathcal{R}}}\bra{e_{\mathcal{L}} g_{\mathcal{R}}} + \hbar\omega_{\mathcal{R}} \ket{g_{\mathcal{L}}e_{\mathcal{R}}}\bra{g_{\mathcal{L}}e_{\mathcal{R}}} +\hbar(\omega_{\mathcal{L}} + \omega_{\mathcal{R}}) \ket{e_\mathcal{L}e_\mathcal{R}} \bra{e_\mathcal{L}e_\mathcal{R}} + J\ket{e_\mathcal{L} g_\mathcal{R}}\bra{g_\mathcal{L}e_\mathcal{R}} + h.c..
\end{equation}
Here $\ket{e_{\mathcal{L}}g_{\mathcal{R}}}$ denotes a state that is comprised of an excited state on the left qubit and a ground state on the right qubit, etc.
If we assume coupling to the two baths is weak, then it is highly unlikely for the doubly excited state $\ket{e_\mathcal{L} e_\mathcal{R}}$ to be populated and it can therefore be neglected.
This gives us a three level model consisting of the states $\lbrace \ket{g_\mathcal{L}g_\mathcal{R}}, \ket{g_\mathcal{L} e_\mathcal{R}}, \ket{e_\mathcal{L}g_\mathcal{R}}\rbrace$ in the so-called site basis.

This resulting three state transport model is isomorphic to the $V$-system model  considered above  \cite{koyu_steady-state_2020,jung_energy_2020}.
In other words, for any $V$-system model we can construct an equivalent two qubit transport model with an appropriate choice of parameters.
The mapping is simple.
The $V$ system model corresponding to the 2-qubit model is obtained by computing the eigenbasis of $\hat{H}_M$ in Eq. (\ref{eq:H2qub}).
This eigenbasis is comprised of the ground state $\ket{g} = \ket{g_\mathcal{L}g_\mathcal{R}}$ and the two singly excited eigenstates:
\begin{subequations}
  \label{eqs:2qubeig}
  \begin{equation}
    \label{eq:eig1}
    \ket{e_1} = -\sin\frac{\theta}{2}\ket{e_\mathcal{L}g_\mathcal{R}} + \cos \frac{\theta}{2}\ket{g_\mathcal{L}e_\mathcal{R}}
  \end{equation}
   \begin{equation}
    \label{eq:eig2}
    \ket{e_2} =  \cos\frac{\theta}{2}\ket{e_\mathcal{L}g_\mathcal{R}} + \sin\frac{\theta}{2}\ket{g_\mathcal{L}e_\mathcal{R}}
  \end{equation}
\end{subequations}
where $\theta=\arctan\left(J/(\Delta_0/2)\right)$  is the mixing angle obtained by diagonalizing Eq. (\ref{eq:H2qub}), $J$ is taken to be real and $\Delta_0 = |\omega_L-\omega_R|/2$ is the site basis energy splitting.
To simplify the algebra, we  consider a symmetric $V$ system model (i.e.,  $\Delta =0$ ), with $p=1$, $r_1 =r =r_2$ and $\gamma_1 = \gamma = \gamma_2$.  (The $\Delta \neq 0$ asymmetric case is treated in Ref. \cite{jung_energy_2020}. )
In this case, the left qubit, which couples to the hot bath simply corresponds to the bright state superposition $\ket{e_1} +\ket{e_2}$ that is excited by the incoherent light, while the right qubit corresponds to the dark state $\ket{e_1} -\ket{e_2}$ that is entirely decoupled from the incoherent light and only couples to the non-radiative cold bath.
Correspondingly, the isomorphism to the two-qubit transport model can be constructed physically by first identifying the transition that is driven by the incoherent light with a hot qubit through which energy flows into the system.
The orthogonal complement of this bright state is a dark state that is not excited by the incoherent light and therefore can only dissipate energy into the cold phonon bath, and hence can be associated with the right qubit through which energy flows out of the system.
The unitary dynamics within the excited state manifold is then responsible for driving the flow of energy between these two transitions, thereby generating a net flow of energy between the two baths.

The particular benefit of the two qubit model is that it allows for a straightforward definition of energy currents that characterize the net energy flux through the system.
Since each qubit is coupled to its own bath and to the other qubit in a linear architecture, the flow of energy through the system can be directly tracked.
This allows for the definition of three different energy currents, from the left bath into the left qubit $\mathcal{J}_{\mathcal{B}_\mathcal{L} \to \mathcal{L}}$, from the left qubit to the right qubit $\mathcal{J}_{\mathcal{L} \to\mathcal{R}}$ and from the right qubit to the right bath $\mathcal{J}_{\mathcal{R} \to\mathcal{B}_\mathcal{R}}$.
Conservation of energy and the linear architecture of these systems indicates that any net energy that flows into the left qubit from the left bath must then flow into the right qubit and then into the right bath, giving the relationship $\mathcal{J} = \mathcal{J}_{\mathcal{B}_\mathcal{L} \to \mathcal{L}} = \mathcal{J}_{\mathcal{L} \to\mathcal{R}} = \mathcal{J}_{\mathcal{R} \to\mathcal{B}_\mathcal{R}}$.
Finally, we can compute these energy currents in the two-qubit model and relate it to the non-equilibrium steady state of the $V$-system model to give:
\begin{equation}
  \label{eq:currents}
  \mathcal{J} = -i \braket{[\hat{\sigma}_L^{z}, \hat{H}_S]} = -2iJ\Tr\lbrace \rho(\ket{e_2}\bra{e_1} - \ket{e_1}\bra{e_2})\rbrace =4J\rho_{e_1e_2}^I,
\end{equation}
where $\hat{\sigma}_L^z= \ket{e_\mathcal{L}}\bra{e_\mathcal{L}} - \ket{g_\mathcal{L}}\bra{g_\mathcal{L}}$ is the Pauli $z$ operator of the left qubit and $\rho_{e_1e_2}^I$ denotes the imaginary part of the coherence between energy eigenstates.
Equation (\ref{eq:currents}) then provides a direct connection between the transport of energy through a quantum system and the coherence between its eigenstates in the non-equilibrium steady state.  Some aspects of Eq. (\ref{eq:currents}) are discussed further in Appendix A of Ref. \cite{jung_energy_2020} and in \cite{caoflux} 

\section{Sample Systems}
\label{sec:examples}

Consider now two physical systems that show these transport-induced coherences in their non-equilibrium steady state.
First, we  discuss the excitation of atomic Calcium by a polarized beam of incoherent light - a simple model system that is a promising candidate for experimental studies of noise-induced coherence.
Then, we consider the consequences of noise-induced coherence on photoinduced biological processes through the example of retinal photoisomerization in sunlight - the first step of human vision.
In both cases, we  examine the predicted behavior of the system under non-secular conditions where noise-induced coherences are properly accounted for, and contrast it to secular dynamics where coherences are assumed to be decoupled from population dynamics.
Comparing the secular and non-secular regimes  highlights the ways in which accounting for noise-induced coherences leads to observably different physical behavior, including modifications of the fluorescence spectrum or  the quantum yield of the photoisomerization process.

\subsection{Calcium Atoms in Polarized Light}
\label{sec:Ca}

\subsubsection{Polarized Partial-Secular Bloch-Redfield Equations}
\label{sec:pol-psbr}

\begin{figure}[ht]
  \centering
  \includegraphics[width=\textwidth]{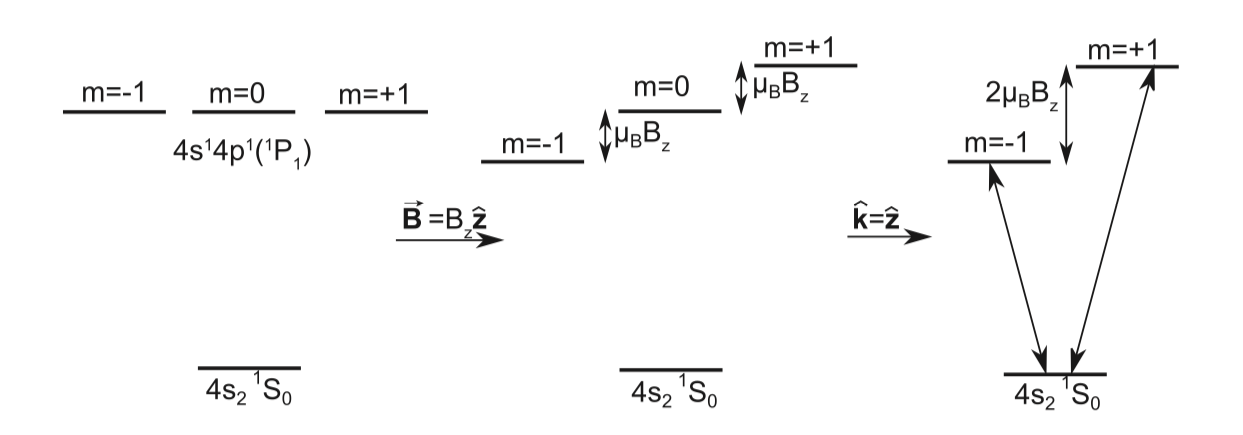}
  \caption{Sketch of the V subsystem of calcium excited by an incident light beam along the $\bm{k} || \hat{z}$ axis. A magnetic field, $\bm{B}$ parallel to the incident light leads to excited-state Zeeman splitting of $2\mu_B B$ between the $m =\pm 1$ states. The rightmost sketch shows the $m=-1$ and $m=+1$ levels, denoted $\ket{e_1}$ and $\ket{e_2}$ henceforth. From Ref. \cite{dodin_secular_2018}.}
  \label{fig:Ca-schem}
\end{figure}

Consider a gas-phase Calcium atom excited by a polarized beam of incoherent light in the presence of a magnetic field.
The ground state of the calcium atom,  $\ket{g} =$$^1S_0$, has both valence electrons in the $4s$ orbital.
Exciting one of the valence electrons into a $4p$ orbital produces a manifold of three $^1P_1$ excited states differentiated by their $m = \{ 0, \pm 1\}$ angular momentum quantum number.
In the absence of a magnetic field, all three excited states are degenerate and separated from the ground state by energy $\hbar \omega_0$
When a magnetic field, $\bm{B} = B_z \hat{\bm{z}}$, is applied along the $z$ axis, the degeneracy between the $m$ states is broken by a Zeeman splitting that shifts the $m=\pm 1$ states by  $\Delta_Z(m) = m \mu_B B_z$, where $\mu_B$ is the magnetic dipole moment of the $4p^{\pm}$ states.
Exciting the calcium atom by a light beam along the $z$ axis causes transitions to the $m=\pm 1$ states, decoupling the $m=0$ state and leading to an effective 3-level $V$-system model with excited states $\ket{e_1} =$ $^1P_1(m=-1)$ and $\ket{e_2} =$ $^1P_1 (m=+1)$, shown schematically in Fig. \ref{fig:Ca-schem}.
The three levels involved in this excitation can then be described by a V-System Hamiltonian.  

\begin{equation}
  \label{eq:Ca-Ham}
  \hat{H}_M = (\hbar \omega_ 0 - \Delta_Z) \ket{e_1}\bra{e_1} + \hbar\omega_0 + \Delta_Z\ket{e_2}\bra{e_2}.
\end{equation}

The incoherent excitation of calcium can be treated using the PSBR equations for a V-system, Eq. (\ref{eqs:psbrme}).
First, consider excitation by isotropic unpolarized light.
The transition dipole moments from the ground to the two $m={\pm 1}$ excited states are orthogonal.
As a result, the $p$ alignment factor vanishes under isotropic excitation, thereby decoupling the populations and coherences.
Consequently, under isotropic excitation the secular and non-secular equations of motion are equivalent.
In contrast, excitation by a specific polarization of incoherent light does generate noise-induced coherence.
The PSBR equations can be rederived in the case of polarized excitation of Calcium to give the following equations of motion:
\begin{subequations}
  \label{eqs:Ca-psbrme}
  \begin{equation}
    \label{eq:Ca-rhoeiei}
    \dot{\rho}_{e_ie_i} = r_i \rho_{gg} - (r_i + \gamma_i)\rho_{e_ie_i} -  \sqrt{r_1r_2} \rho_{e_1e_2}^R
  \end{equation}
  \begin{align}
    \label{eq:Ca-rhoe1e2}
    \dot{\rho}_{e_1e_2} =& \sqrt{r_1r_2}\rho_{gg} -i\Delta \rho_{e_1e_2} -\half(r_1 + r_2 + \gamma_1 + \gamma_2)\rho_{e_1e_2}\\
                          &- \frac{1}{2}\sqrt{r_1r_2}( \rho_{e_1e_1} + \rho_{e_2e_2}),
  \end{align}
\end{subequations}
where the $\gamma_i$'s are defined as in Eq. (\ref{eqs:psbrme}) but the pumping rates in the polarized case $r_i = 3 r_i^{(\mathrm{iso})}/16\pi$ are smaller than the isotropic pumping rates, $r_i^{(\mathrm{iso})}$, since only one polarization mode of the driving field is occupied.
For the excitation of calcium atoms the excitation and spontaneous emission rates are the same for both excited states $r_1 = r = r_2$ and $\gamma_1 = \gamma = \gamma_2$.

Contrasting the polarized equation of motion, Eq. (\ref{eqs:Ca-psbrme}) with the isotropic case in Eq. (\ref{eqs:psbrme}) shows that the excitation and spontaneous emission pathways have different $p$ alignment values.
The polarized exciting field leads to the case of $p=1$ with full noise-induced coherence generated by the exciting light (i.e. the excitation pathways are able to interfere).
By contrast, since spontaneous emission is driven by interaction with the isotropic vacuum field, the orthogonal transition dipole moments of the two excited states leads to no net interference in the spontaneous emission pathways,
we see this in the vanishing $\sim\sqrt{\gamma_1\gamma_2}$ term which describes interference in the spontaneous emission of the two excited states. 

\subsubsection{Dynamics \& Atomic Emission Spectra}
\label{sec:ca-spec}

\begin{figure}[ht]
  \centering
  \includegraphics[width=\textwidth]{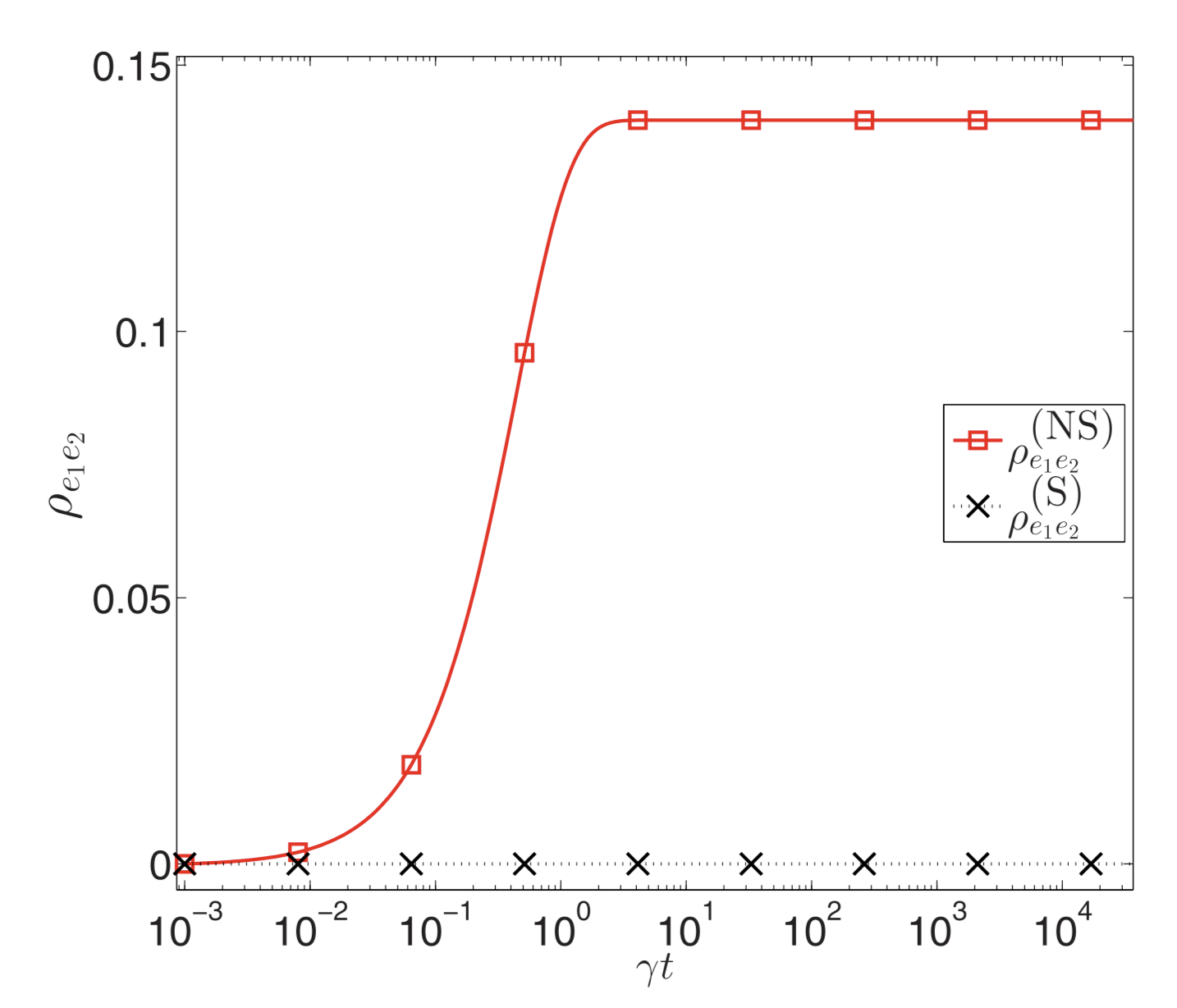}
  \caption{Coherences of a calcium atom in the small splitting regime $\Delta = 0.012 \gamma$ irradiated by polarized light from a blackbody source at $T = 5800 K$ whose average photon occupation number at the transition energies is $\overline{n}=0.0633$. The natural line width of calcium is $\gamma = 2\pi \times 34.6$ MHz. The non-secular solution is indicated by red squares and secular solution by black $\times$'s.  Reproduced with permission from Ref. \cite{dodin_secular_2018}.}
  \label{fig:Ca-dyn}
\end{figure}

The similarity between the polarized and isotropic PSBR master equations leads to very similar dynamics in the two cases.
By applying the same techniques used to solve the isotropic case, analytical dynamics can also be obtained for polarized excitation \cite{dodin_secular_2018} with similar underdamped and overdamped regimes as shown in Figs. \ref{amrslowfig:2} and \ref{amrslowfig:3}.
In order to highlight the experimental consequences of noise-induced coherences we focus on the overdamped regime where the coherences are of comparable magnitude to the eigenstate populations.
By solving the fully non-secular Eq. (\ref{eqs:Ca-psbrme}) as well as its secular approximation where populations and coherences are decoupled, we obtain the following dynamics in the overdamped $\Delta/\gamma \ll 1$:
\begin{subequations}
  \label{eqs:Ca-dyn}
  \begin{equation}
    \label{eq:Ca-pops}
    \rho_{e_ie_i}(t) = \frac{r}{\gamma}(1-e^{-\gamma t}),
  \end{equation}
  \begin{equation}
    \label{eq:Ca-sec}
    \rho_{e_1e_2}^{(S)}(t) = 0,
  \end{equation}
  \begin{equation}
    \label{eq:Ca-ns}
    \rho_{e_1e_2}^{(NS)}(t) = \rho_{e_ie_i}(t),
  \end{equation}
\end{subequations}
where $\rho_{e_1e_2}^{(S)}$ denotes the coherences in the secular approximation and $\rho_{e_1e_2}^{(NS)}$ are the coherences in the full non-secular solution.
As discussed for the isotropic case in Fig. \ref{amrslowfig:2}, the coherences in the non-secular case are small enough that they do not modify population dynamics, leading to excited state populations equivalent to the secular case.
Nevertheless, the coherences themselves do not vanish and remain finite.
These solutions are shown for parameters consistent for a calcium atom in a weak magnetic field in Fig. \ref{fig:Ca-dyn}.
We note that in this case, the non-secular steady-state of the calcium atom shows coherences between energy eigenstates.
These steady-state coherences are transport-induced coherences that arise due to transport from the high-temperature bath of the polarized exciting light and the low-temperature isotropic vacuum modes.  The secular result is shown for comparison, and its range of validity is discussed in Ref. \cite{dodin_secular_2018}.

Although the eigenstate populations are not affected by coupling to the coherences in the non-secular regime, the presence of these coherences can change the directionality of excited state emission, allowing them to be observed.
In particular, the intensity of light emitted from a calcium atom at a point $\bm{R} = (R, \theta, \phi)$ away from it is given by
\begin{equation}
  \label{eq:Ca-Intensity}
  \langle I (\bm{R}, t)\rangle = I_0\left[ \frac{1 + \cos^2\theta}{2} \left[ \rho_{e_1e_1}(t') + \rho_{e_2 e_2}(t')\right]
    +\sin^2\theta\left( \rho_{e_1e_2}^R(t')\cos 2\phi -\rho_{e_1e_2}^I(t')\sin2\phi\right)\right],
\end{equation}
where $t' = t +R/c$, $I_0 =  {\overline{n} \omega_0^4}/{32\pi^2 \epsilon_0 c^3 R^2}$, $\epsilon_0$ is the permitivity of free space and $c$ is the speed of light.

The influence of coherence on the spatial emission profile allows for the direct observation of noise-induced coherences and their dynamics.
One possible detection scheme compares  emission integrated over certain subsets of the sphere.
Define the following subregions of the sphere:
(a) detection integrated over all directions, denoted by $I_z$,
(b) collecting the light in the two quarter spheres with $ \theta \in [0, \pi]$ and $\phi \in[-\pi/4, \pi/4] \cup [3\pi/4, 5\pi/4]$ denoted $I_A$, and
(c) collecting the light in the two quarter-spheres with $\theta \in [0, \pi]$ and $\phi \in [0, \pi/2] \cup [\pi, 3\pi/2]$ denoted $I_B$.
This allows us to define the following detection schemes for the populations and coherences
\begin{subequations}
  \label{eqs:Ca-detect}
  \begin{equation}
    \label{eq:Ca-I-pop}
    I_z = \frac{8\pi}{3} I_0 \left(\rho_{e_1e_1} + \rho_{e_2e_2}\right)
  \end{equation}
  \begin{equation}
    \label{eq:Ca-I-real}
    I_A - I_A' = \frac{16}{3}I_0 \rho_{e_1e_2}^R
  \end{equation}
  \begin{equation}
    \label{eq:Ca-I-imag}
    I_B - I_b' = -\frac{16}{3} I_0 \rho_{e_1e_2}^I
  \end{equation}
\end{subequations}
where $I_A'  =I_z- I_A$ and $I_B' = I_z - I_B$ is the intensity in the regions complementary to $I_A$ and $I_B$ respectively.

\subsection{Photoisomerization of Retinal in Sunlight}
\label{sec:Retinal}
\begin{figure}[htbp]
  \centering
  \includegraphics[width=\textwidth]{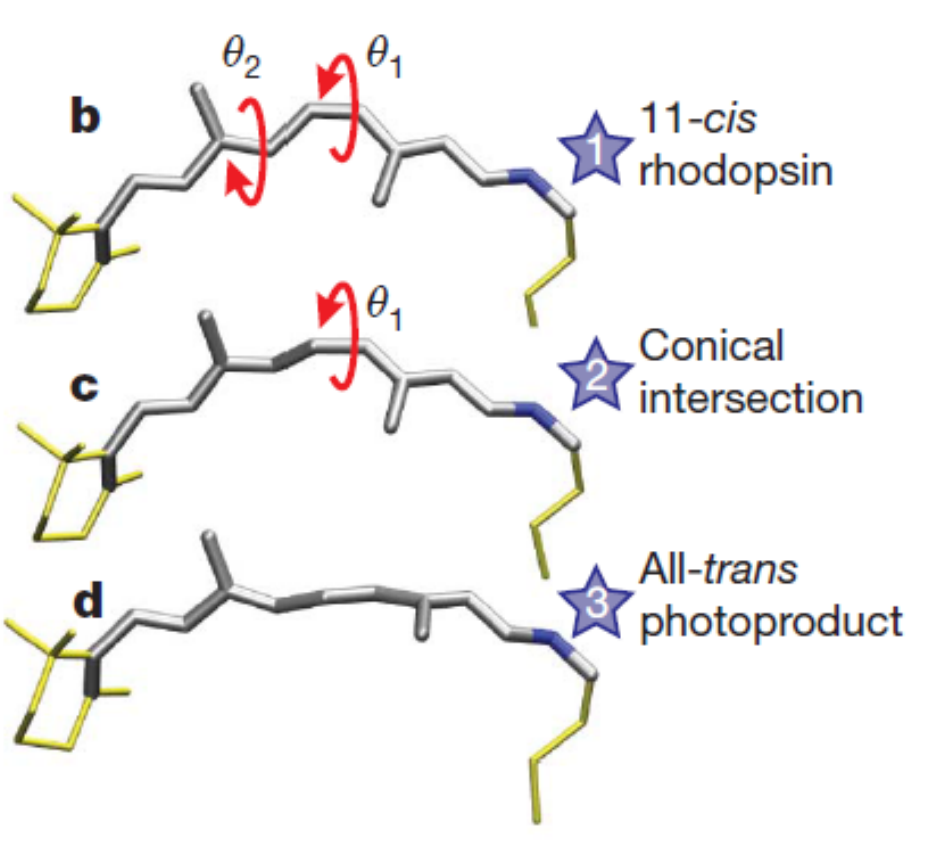}
  \caption{Schematic representation of the cis-trans photoisomerization of retinal, the first step in human vision.  Given time-dependent excitation, the initial 11-cis-retinal structure (1) is photo-excited, then evolves  through a conical intersection (2) into the all-trans photoproduct (3). The process takes place within a complex molecular environment (not shown).  From Ref. \cite{polli_conical_2010}.}
  \label{fig:retinal-schem}
\end{figure}

We now consider noise-induced coherences in a more complex biomolecular system, the photoisomerization of retinal shown in Fig. \ref{fig:retinal-schem}.  This process, embedded in a complex molecular environment,  comprises the first step in human vision.  That is, light incident on the eye induces isomerization of the retinal, as discussed below.  This generates a sequence of chemical reactions, ending with nerve signals to the brain.
Capturing noise-induced coherence in photoinduced biomolecular processes is crucial to understanding their dynamics in nature.
While experimental studies, most often based on laser excitation, provide important insight into the properties of these systems, they fail to capture the process in realistic situations where noisy incoherent light is their primary driver.
In this section, we examine the dynamics of retinal photoisomerization under continuous driving by incoherent light using a non-secular approach that captures noise-induced coherence, and a commonly used secular rate-law equation that decouples coherences and populations.
We will see how the presence of noise-induced coherences generated by the incoherent light modifies the non-radiative relaxation pathways through interference between energy eigenstates.
As a result, the excited state predicted by properly accounting for noise-induced coherences relaxes differently into the ground state, modifying the quantum yield of the photoisomerization in comparison to the secular rate-law approximation.
Moreover, the transport of energy from the hot bath of incoherent light into the cold phonon bath that is responsible for non-radiative relaxation establishes transport-induced coherences between energy eigenstates.
These observations are particularly important in the search for quantum effects in biology as they show that if such effects are present under realistic conditions they will appear very different in form and mechanism from those observed in laser driven experiments.  Our analysis stops short of examining the steady state itself.  Rather, the focus is on the way in which noise induced coherences are generated in the system when initiated with light of differing turn-on times.  (For a steady-state study of an alternate light-induced biological process, energy transfer in LH1-RC, see \cite{chuang_lh1rc_2020}.)

\subsubsection{The Two-State Two-Mode Model}
\label{sec:2S2M}
We  describe the dynamics of retinal in rhodopsin by combining the PSBR equation \cite{tscherbul_partial_2015} for incoherent excitation with the widely used Two-State Two-Mode (2S2M) model \cite{hahn_femtosecond_2000,hahn_quantum-mechanical_2000} for the system and bath dynamics.
The 2S2M model is a minimal vibronic model of retinal that includes the key electronic and nuclear degrees of freedom and captures the observed quantum yield and transient absorption spectra on the picosecond time scale \cite{tscherbul_excitation_2014,balzer_modeling_2005}.
It considers two electronic states, a ground $\ket{0}$ and excited state $\ket{1}$, with potential energy surfaces that describe two collective nuclear modes -  a low frequency torsional mode $\phi \in [-\pi/2, 3\pi/2]$ whose rotation  captures the cis-trans isomerization, and a high-frequency stretch mode $x$ that represents other nuclear motions in the molecule.

\begin{figure}[htbp]
  \centering
  \includegraphics[width=\textwidth]{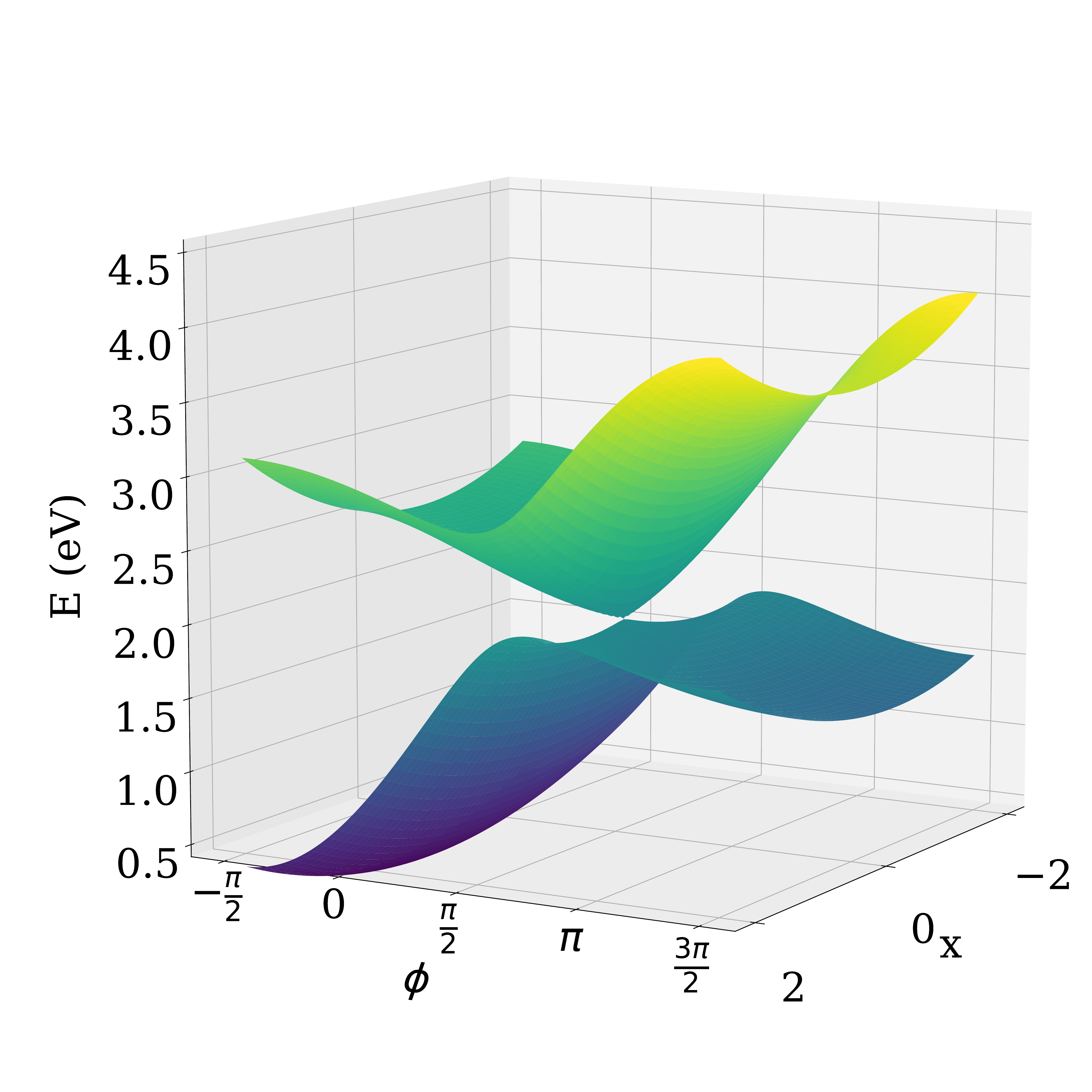}
  \caption{Adiabatic potential energy surfaces of the two-state two-mode model of retinal described by Eq. (\ref{eq:2s2M-Ham}). The adiabatic curves show a conical intersection at $x=0$ and $\phi = \pi/2$. The 11-cis-isomer corresponds to the region $\phi \in [-\pi/2, \pi/2]$ and the all-trans-isomer to the complementary region $\phi \in [\pi/2, 3\pi/2]$. From Ref. \cite{dodin_light-induced_2019}.}
  \label{fig:retinal-PES}
\end{figure}

The 2S2M model represents the system by a molecular Hamiltonian of the form
\begin{equation}
  \label{eq:2s2M-Ham}
  \hat{H}_M = \sum^1_{n,n'=0}\ket{n}\bra{n'}\left[\delta_{n,n'}\left(\hat{T} +E_n +(-1)^n \half\tilde{V}_n(1-\cos\phi)+\half\omega x^2 +\kappa x \delta_{n,1}\right) +(1-\delta_{n,n'})\lambda x\right]
\end{equation}
where $\hat{T}$ is the nuclear kinetic energy operator, $E_n$ are the electronic state energies and $\kappa, \omega, \lambda$ and $\bar V_n$ are the 2S2M model parameters.
We use the parameter values computed in Refs. \cite{hahn_femtosecond_2000} and \cite{hahn_quantum-mechanical_2000}.  (The need for improved parameters is discussed in \cite{vargas-hernandez_multi-objective_nodate} but this does alter our discussion.)
This Hamiltonian produces a conical intersection at $x=0$ and $\phi=\pi/2$ that  leads to rapid non-radiative decay from the excited state \cite{balzer_mechanism_2003}.
The adiabatic potential energy surfaces are shown in Fig. \ref{fig:retinal-PES}.

The density matrix dynamics can then be expressed in the following form:
\begin{subequations}
  \label{eqs:retinal-EOM}
  \begin{equation}
    \label{eq:retinal-ME}
    \hat{\dot{\rho}}(t) = \left(\hat{\hat{\mathcal{L}}}_0 + \hat{\hat{\mathcal{L}}}_{\mathrm{rad}} +  \hat{\hat{\mathcal{L}}}_{\mathrm{phon}}\right)\hat{\rho}(t)
  \end{equation}
  \begin{equation}
    \label{eq:retinal-L_0}
    \hat{\hat{\mathcal{L}}}_0 \hat{\rho}(t) = -\frac{i}{\hbar}\left[ \hat{H}_M, \hat{\rho}\right]
  \end{equation}
\end{subequations}
where $\hat{\hat{\mathcal{L}}}_0$ is the unitary dynamics under Hamiltonian $\hat{H}_M$ defined in Eq. (\ref{eq:2s2M-Ham}), $\hat{\hat{\mathcal{L}}}_{\mathrm{rad}}$ encodes the driving by the incoherent radiative field, and $\hat{\hat{\mathcal{L}}}_{\mathrm{phon}}$ encodes the dynamics induced by the phonon bath.
We  model the radiative coupling using both the PSBR equations as well as the secular rate laws in order to examine the effect of properly including noise-induced  coherence.
The phonon-induced dynamics can then be treated using the Markovian Redfield equations.
The details and derivations of these non-unitary dynamics can be found in Ref. \cite{dodin_light-induced_2019}.

\subsubsection{State Dynamics \& Photoisomerization Quantum Yield}
\label{sec:Ret-QY}

To characterize noise-induced coherence in these systems, Eq. (\ref{eqs:retinal-EOM}) can be numerically integrated in the vibronic eigenbasis of the 2S2M Hamiltonian, Eq. (\ref{eq:2s2M-Ham}). We perform a PSBR simulation explicitly propagating 900 system eigenstates.  
We perform this numerical integration for both the non-secular and secular rate-law approximation to the incoherent excitation and for a variety of turn-on times $\tau_r$ of the incoherent light. The initial excitation of the system is followed by relaxation to lower energy states.
By comparing the secular and non-secular results we are able to expose the interference that arises due to the initially prepared noise-induced coherence on the non-radiative relaxation pathways through the energy eigenbasis.
To quantify this behavior we display  the populations and coherences between several representative pairs of states,  one pair each in the bright state manifold of excited states that are initially excited by the light, one pair of dark intermediates states through which the system undergoes non-radiative relaxation, and one pair of dark product states in the ground state of the system.
Since noise-induced coherences are bounded by the population of the pair of states, the value of coherence  captures both how well-defined the relative phase is between two states as well as their populations.
In order to separately quantify  the degree of coherence, we  also consider the coherence ratio,
\begin{equation}
  \label{eq:Coh-Ratio}
  C_{ij}(t) = \frac{|\rho_{ij}(t)|^2}{\rho_{ii}(t) \rho_{jj}(t)},
\end{equation}
which normalizes out the effect of state population to give a number $C_{ij} \in [0, 1]$.
This measure has a value of unity when the quantum system is in a coherent superposition of the two states, and monotonically decays to zero for a fully incoherent mixture, irrespective of the state populations.

\begin{figure}[htbp]
  \centering
  \includegraphics[width=\textwidth]{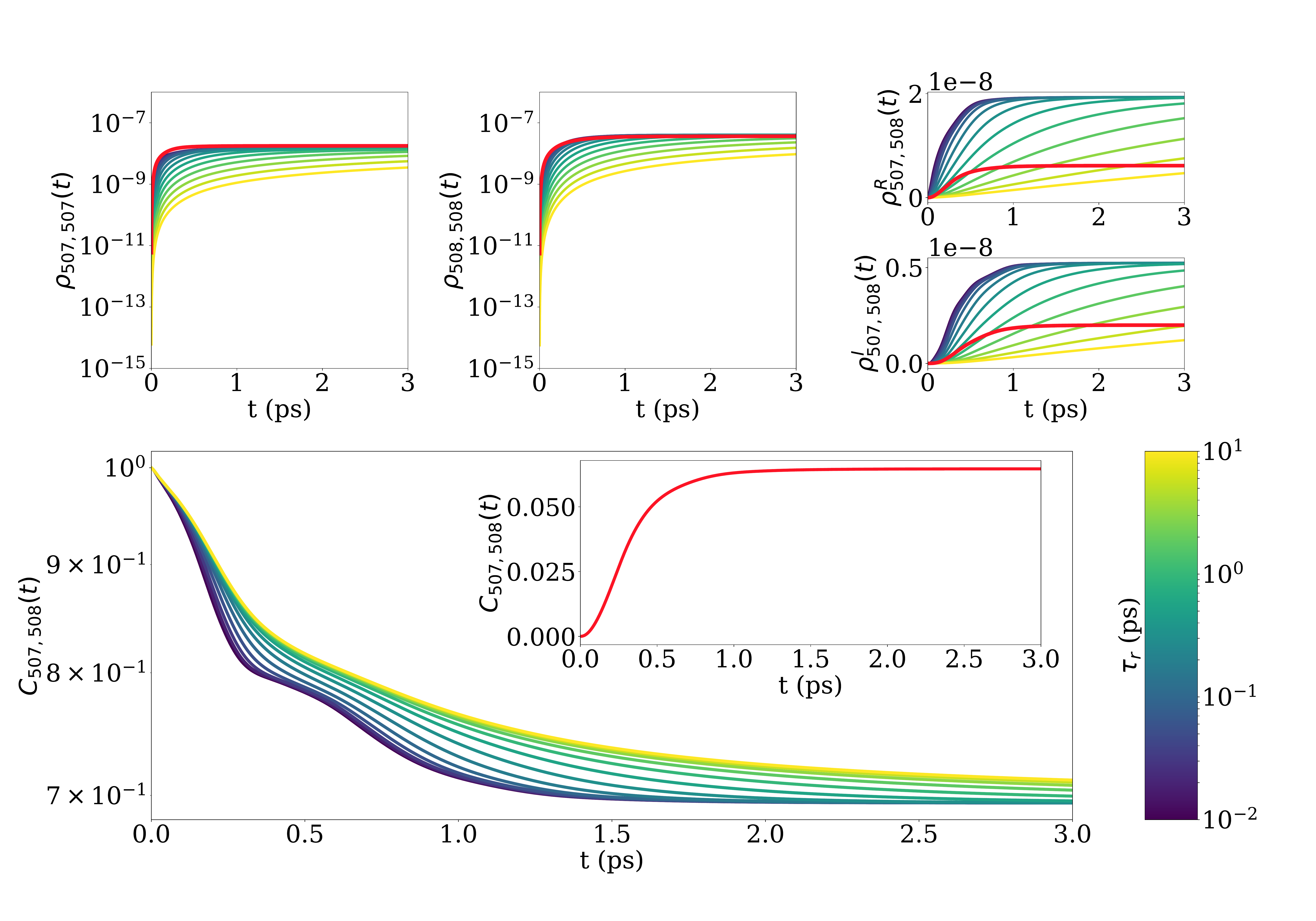}
  \caption{Density matrix elements (top) and coherence ratio (bottom) corresponding to a pair of nearly degenerate bright excited states $\ket{507}$ and $\ket{508}$ under a range of light-field turn-on times (yellow-blue traces). These are contrasted with a secular approximation of the excitation (red trace) where the excitation generates no coherences. Reproduced from Ref. \cite{dodin_light-induced_2019}.}
  \label{fig:retinal-bright}
\end{figure}

Consider first a pair of bright excited states, shown in Fig. \ref{fig:retinal-bright}, that are initially excited by  incoherent light with various turn-on times $\tau_r$.
By examining the coherence ratio, we see that for all turn-on times, non-secular excitation initially prepares a coherent superposition of these two states that eventually loses some of its coherent character over  a few picoseconds, to reach an apparently coherent steady state with $C_{ij} \approx 0.7$.
By contrast, the secular excitation initially prepares an incoherent mixture of the two states with $C_{ij}(0) = 0$ but after a few picoseconds shows an increase in its coherence ratio reaching a small apparent steady-state value of $C_{ij} = 0.05$.  
This coherence was not generated by the incoherent light but rather by interference in the non-radiative relaxation pathways through a mechanism similar to vacuum-induced coherence in quantum optics \cite{agarwal_quantum_1974,agarwal_quantum_2012,ficek_quantum_2005}.  
Interference in the non-radiative relaxation will lead to different decay rates for different phases of excited state superpositions.
A statistical mixture of excitations with randomly assigned phase will then see some superpositions decay more rapidly due to constructive interference in non-radiative relaxation whereas others will decay more slowly due to destructive interference.
This leads to a disproportionate survival of destructively interfering excitations and the emergence of the decay-induced coherences seen in the red traces of Fig. \ref{fig:retinal-bright}.

\begin{figure}[htbp]
  \centering
  \includegraphics[width=\textwidth]{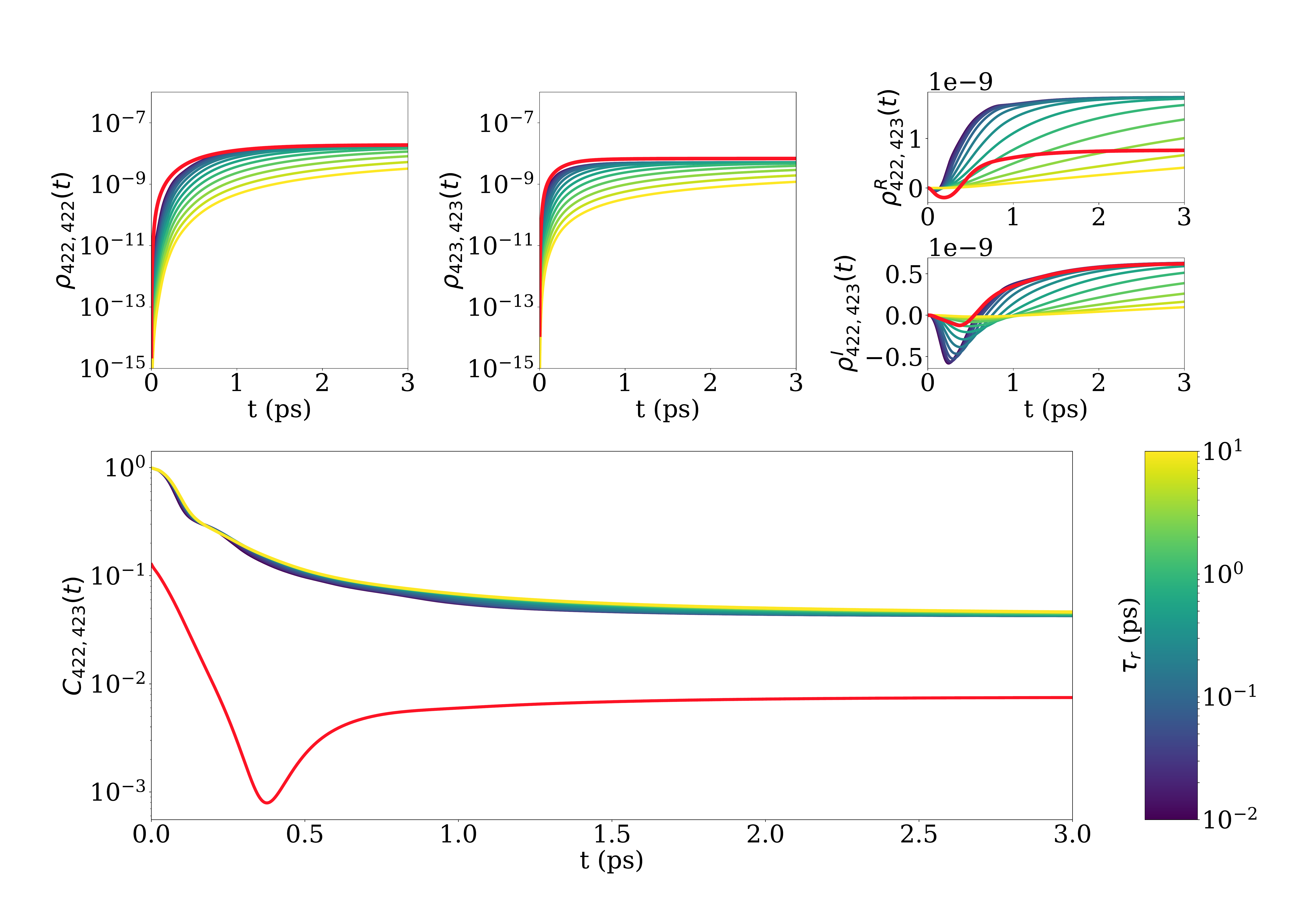}
  \caption{The density matrix elements (top) and coherence ratio (bottom) corresponding to a pair of dark intermediate states $\ket{422}$ and $\ket{423}$ under a range of light-field turn-on times (yellow-blue traces). These are contrasted with a secular approximation of the excitation (red trace) where the excitation generates no coherences. Reproduced from Ref. \cite{dodin_light-induced_2019}.}
  \label{fig:retinal-int}
\end{figure}

\begin{figure}[htbp]
  \centering
  \includegraphics[width=\textwidth]{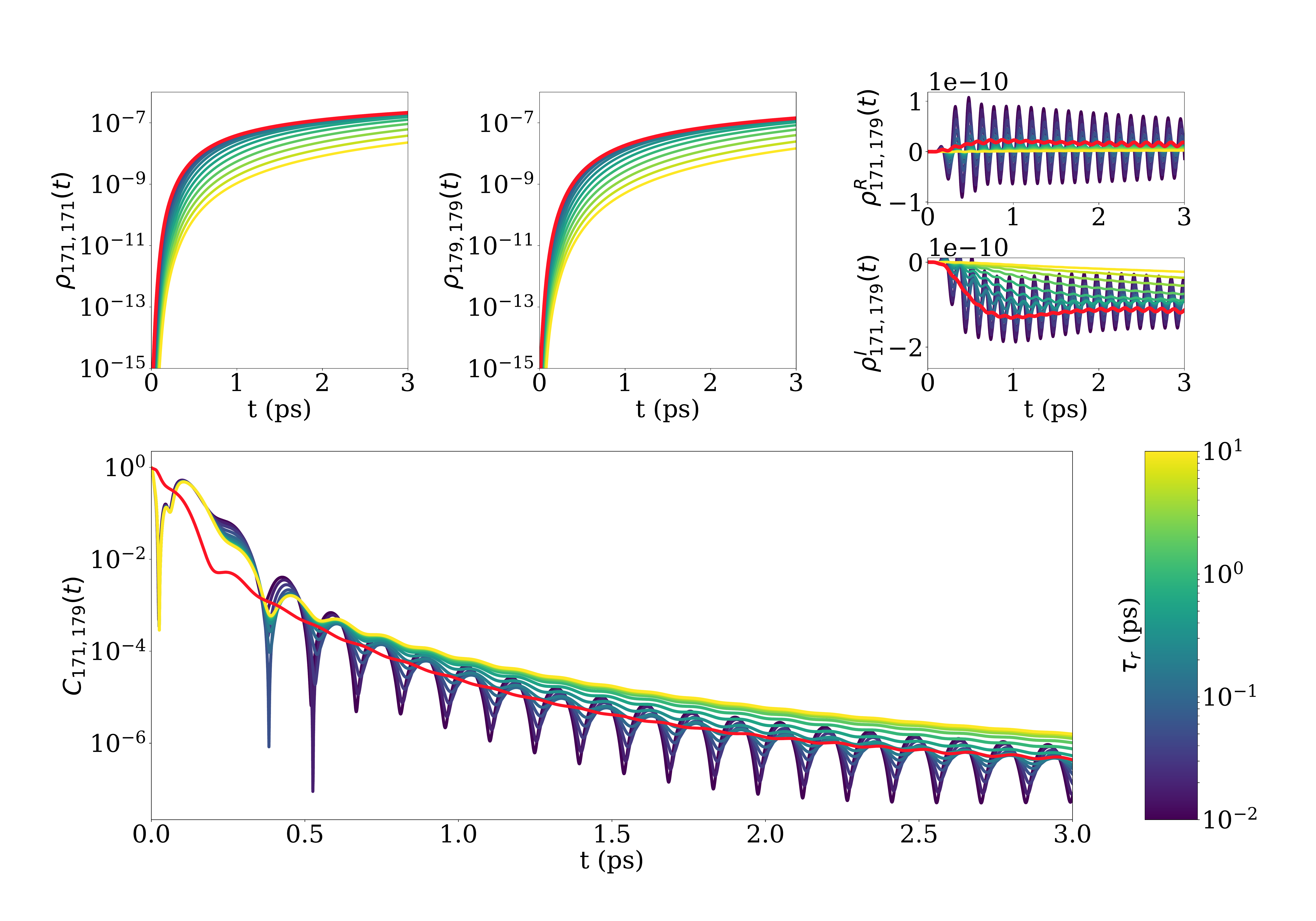}
  \caption{The density matrix elements (top) and coherence ratio (bottom) corresponding to a pair of dark product states $\ket{171}$ and $\ket{179}$ under a range of light-field turn-on times (yellow-blue traces). These are contrasted with a secular approximation of the excitation (red trace) where the excitation generates no coherences. Reproduced from Ref. \cite{dodin_light-induced_2019}.}
  \label{fig:retinal-product}
\end{figure}

States initially excited by the incoherent light undergo non-radiative relaxation, producing populations and  coherences in the dark intermediate and product state manifolds in Figs. \ref{fig:retinal-int} and \ref{fig:retinal-product}.  The coherences they display therefore reflect the net effect of interference between the  many non-radiative relaxation pathways that pass through these eigenstates.
Nevertheless, at early times, non-secular excitation leads to fully coherent superpositions between the states with $C_{ij}(0) = 1$.
This indicates that the initial relaxation pathways proceed entirely through coherent superpositions of energy eigenstates with the magnitude of the coherences, and therefore the interference effects, slowly decreasing over time.
Nevertheless, some residual transport induced coherence persists even between these states.
By considering the secular (red) traces we see that secular excitation leads to non-radiative relaxation through different pathways, most clearly visible in the differences between the coherence plots for secular and non-secular excitation.
However, the initial strength of noise-induced coherence increases for states further down the relaxation pathways, with the product state showing a maximal initial coherence of $C_{ij}(0)=1$.
As a result, the initial population of the product state-manifold is dominated by the specific superpositions that showed the maximal constructive interference in the non-radiative pathways connecting the bright excited states to the dark product states.

Since this non-radiative relaxation dynamics can be highly complex, involving interfering pathways through hundreds of energy eigenstates, it is difficult to identify the overall effect of this interference on the photoisomerization pathways.
In addition to these state-specific measures of noise-induced coherence, it is therefore useful to  examine the global effect of noise-induced coherence by considering the instantaneous quantum yield of the photoisomerization process
\begin{subequations}
  \label{eqs:Ret-QY}
  \begin{equation}
    \label{eq:Ret-QY}
    Y_1 (t) = \frac{\braket{\hat{P}_{\mathrm{trans}}^{(1)}(t)}}{\braket{\hat{P}_{\mathrm{cis}}^{(0)}} + \braket{\hat{P}_{\mathrm{trans}}^{(1)}(t)}}
  \end{equation}
  \begin{equation}
    \label{eq:Ret-Proj_cis}
    \hat{P}_{\mathrm{cis}}^{(0)} = \Theta(\pi/2- |\phi|)\ket{0}\bra{0}
  \end{equation}
  \begin{equation}
    \label{eq:Ret-Proj_trans}
    \hat{P}_{\mathrm{trans}}^{(1)} = \Theta(|\phi|- \pi/2)\ket{1}\bra{1}
  \end{equation}
\end{subequations}
where the cis and trans projection operators  $\hat{P}_{\mathrm{cis}}^{(0)}$ and  $\hat{P}_{\mathrm{trans}}^{(1)}$ are taken over the $\ket{0}$ and $\ket{1}$ diabatic surfaces respectively.
This measure captures the overall effect of coherence on the relaxation process and provides complementary information to the state-specific measures discussed above.

\begin{figure}[htbp]
  \centering
  \includegraphics[width=\textwidth]{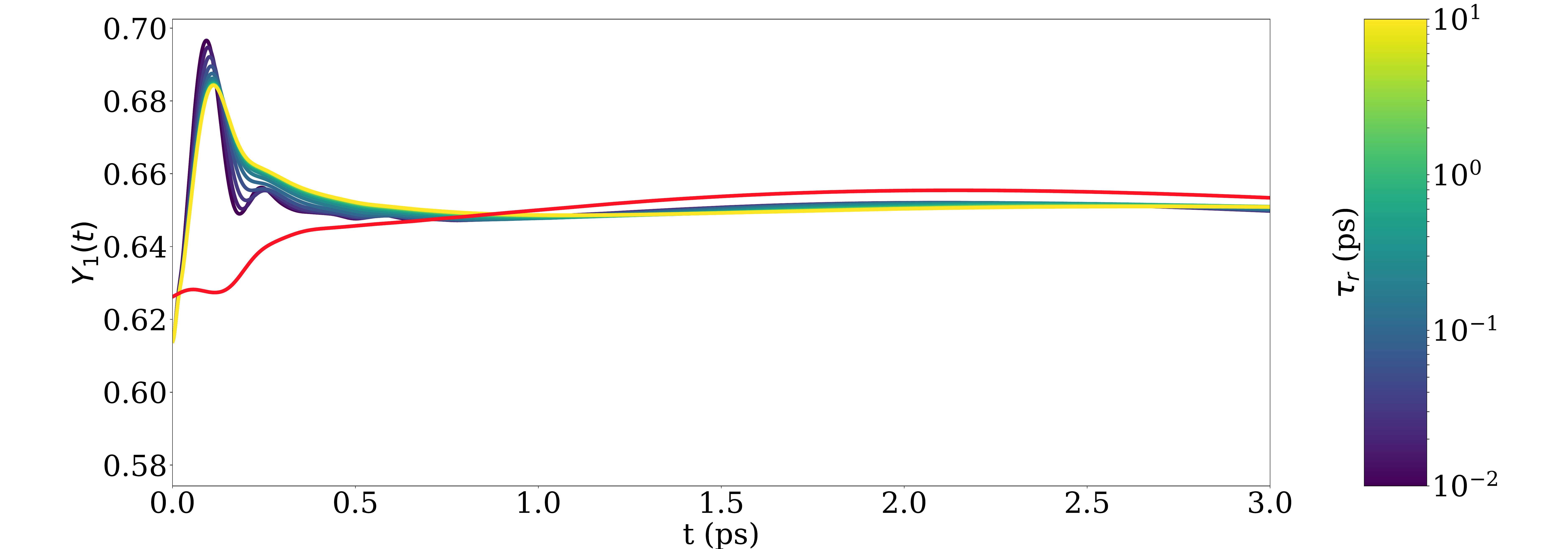}
  \caption{Time-dependent quantum yield with varying turn-on times $\tau_r$ (yellow-blue traces). These are contrasted with a secular approximation to the incoherent excitation (red).  Reproduced from Ref. \cite{dodin_light-induced_2019}.}
  \label{fig:retinal-qy}
\end{figure}

By examining the time-dependent quantum yield in Fig. \ref{fig:retinal-qy}, we see that the differences in the initially excited bright states can lead to different quantum yields, indicating that interference due to noise-induced coherence can impact nuclear dynamics in biomolecular systems.
While this effect  is most pronounced in the first 500 femtoseconds, we see a residual difference between secular and non-secular excitation even on longer picosecond timescales.  Even longer timescales show interesting effects on the way to the steady state, as discussed in Ref. \cite{axelrod_multiple_2019}.

\section{Conclusion}
\label{sec:conc}
To summarize, in this tutorial we've shown that since the stochastic nature of incoherent light differs significantly from laser sources in its temporal characteristics so does the molecular excitations it generates.
Surprisingly, in spite of the lack of phase information in incoherent light, simultaneous excitation from the ground state to two different excited states can still preferentially populate  excited states with coherence,  determined by the geometry of the molecular or atomic system.
The resulting coherence generated by this process can subsequently lead to interference in the relaxation dynamics of the system.
These coherences will generally  vanish on long time scales if the system is driven by a single bath as the system relaxes towards equilibrium.
However, if the system is coupled to two baths at different temperatures, the effect of the two  baths on the system generates coherences in the non-equilibrium steady state.
In addition to the concepts covered in this tutorial, a number of related topics continue to be active areas of ongoing research.

\textit{(1) Slow Turn-on \& Generalized Adiabatic Dynamics:}
The loss of transient dynamics has been consistently observed when studying incoherent excitation by slowly turned-on incoherent light.
This feature is not a peculiarity of these models or of incoherent excitation but is in fact a necessary implication of a generalized adiabatic theorem, the Adiabatic Modulation Theorem, that governs the dynamics induced by a potentially rapidly varying field as one of its properties (e.g. intensity) is slowly varied \cite{dodin_generalized_2021}.
In fact, this theorem proves that under realistic illumination where the intensity of the light is slowly turned-on relative to molecular timescales, the only coherences that can appear are those in the steady state.
This observation has motivated the development of new techniques for efficiently computing molecular steady states under incoherent excitation \cite{axelrod_multiple_2019,axelrod_efficient_2018,loaiza_computational_2021}.   

\textit{(2) Experimental Validation of Noise-Induced Coherence:}
To date, nearly all studies of noise-induced coherence have been theoretical and computational investigations.
However, several of these studies predict experimentally observable implications of the interference described by these coherences.
riments have been proposed to verify the impact of noise-induced interference on the fluorescence of atomic
 Calcium excited by incoherent light \cite{dodin_secular_2018,koyu_steady-state_2020}.
In addition, experimental methods for probing noise-induced excitations using existing laser apparatus have also been suggested to enable the study of noise-induced dynamics directly in biological systems \cite{chenu_thermal_2015,chenu_transform-limited-pulse_2016}.  In addition, related experiments have been performed on vacuum induced coherence \cite{agarwal_quantum_2012,ficek_quantum_2005}.

\textit{(3) Relevance to Biological Systems:}
Finally, a number of theoretical efforts are ongoing to identify the impact of noise-induced coherence on the efficiency of biological systems.
For example, several recent studies have examined the effect of different excitation and trapping conditions on the efficiency of model photosynthetic complexes \cite{kassal_does_2013,leon-montiel_importance_2014,jung_energy_2020,chuang_lh1rc_2020,tscherbul_non-equilibrium_2018} and of retinal photoisomerization in human vision \cite{tscherbul_excitation_2014,tscherbul_quantum_2015,dodin_light-induced_2019,chuang_extreme_2020}.
These studies aim to identify circumstances where quantum effects may play role in biological processes under physiological conditions.

\textbf{Acknowledgment}:  This work was supported by the U.S. Air Force Office of Scientific Research (AFOSR) grant FA9550-20-1-0354.

\bibliography{coherences} 

\end{document}